\documentclass{aa}

\usepackage{graphicx}
\usepackage{txfonts}
\usepackage{academicons}
\usepackage{amsmath}

\usepackage{natbib,twoopt}
\usepackage[dvipsnames]{xcolor}

\newcommand{\Gaia}{{\it Gaia}}

\newcommand{\gspspec}{{\it GSP-Spec}}
\newcommand{\Lit}{{\rm Lit.}}

\newcommand{\T}{$T_{\rm eff}$}
\newcommand{\Tm}{T_{\rm eff}}
\newcommand{\g}{log($g$)}

\newcommand{\meta}{[M/H]}
\newcommand{\alfa}{$\alpha$}
\newcommand{\AF}{[\alfa/Fe]}

\newcommand{\Vrad}{$V_{\rm Rad}$}

\newcommand{\Rstar}{$R_{\star}$}
\newcommand{\RstarLitt}{$R_\star^{\rm Lit}$}
\newcommand{\RstarGaia}{$R_\star^{\rm Gaia}$}
\newcommand{\Mstar}{$M_{\star}$}
\newcommand{\MstarLitt}{$M_\star^{\rm Lit}$}
\newcommand{\MstarGaia}{$M_\star^{\rm Gaia}$}

\newcommand{\Rp}{$R_{\rm p}$}
\newcommand{\RpLitt}{$R_{\rm p}^{\rm Lit}$}
\newcommand{\RpGaia}{$R_{\rm p}^{\rm Gaia}$}
\newcommand{\Mp}{$M_{\rm p}$}
\newcommand{\MpLitt}{$M_{\rm p}^{\rm Lit}$}
\newcommand{\MpGaia}{$M_{\rm p}^{\rm Gaia}$}

\newcommand{\Rjup}{$R_{\rm Jup}$}
\newcommand{\Mjup}{$M_{\rm Jup}$}

\newcommand{\SNR}{$S/N$}

\newcommand{\CaFe}{[Ca/Fe]}

\usepackage{listings}
\definecolor{dkgreen}{rgb}{0,0.6,0}
\definecolor{gray}{rgb}{0.5,0.5,0.5}
\definecolor{mauve}{rgb}{0.58,0,0.82}

\begin{document} 

   \title{The \Gaia\ spectroscopic catalogue of exoplanets and host stars}

   \author{
          Patrick de Laverny
          \and 
          Roxanne Ligi
          \and
          Aurélien Crida
          \and     
          Alejandra Recio-Blanco
          \and
          Pedro A. Palicio
          } 

   \institute{
   Université Côte d'Azur, Observatoire de la Côte d'Azur, CNRS, Laboratoire Lagrange, Bd de l'Observatoire, CS 34229, 06304 Nice cedex 4, France  
   }

   \date{Received March 25, 2025 ; accepted May 17, 2025}
   
   \abstract
   {Complete, accurate, and precise catalogues of exoplanet host star (EHS) properties are essential to deriving high-quality exoplanet parameters. These datasets can then be used to study individual planets, planet populations, and planet formation within their Galactic context.}
   {This paper is aimed at homogeneously parameterising EHS and their exoplanets, selected from the Encyclopedia of Exoplanetary Systems and the NASA Exoplanets Archive, using \Gaia\ astrometric, photometric, and \gspspec\ spectroscopic data, complemented by some ground-based spectroscopic survey information.}
   {From the atmospheric parameters of 2573 EHS, we computed their luminosity, radius, and mass, with no prior assumption from stellar evolution models. Their Galactic positions, kinematic and orbital properties were also derived. We then re-scaled the mass and radius of 3556 exoplanets, fully consistently with the stellar data (when available).}
   {The \Gaia\ spectroscopic stellar 
effective temperatures, luminosities, and radii are in rather good
agreement with literature values but are more precise. 
In particular, stellar radii are derived with typically less than 3\% uncertainty (instead of $\sim 8\%$ in the literature); this reduces  the uncertainty on the planetary radii significantly and allows for a finer analysis of the decrease in the number of planets around $1.8\,R_\oplus$ (evaporation valley). 
Larger differences, however, were found for the masses that are more difficult to estimate by any methods. We note that the EHS population is rather diverse in terms of the chemical and Galactic properties, although they are all found in the Solar vicinity, close to the Local spiral arm. Most EHS belong to the thin disc, but some older thick disc and halo members have also been identified. 
For the less massive planets ($\log(M_p/M_{Jup})\lesssim-0.6$), the average planet radius increases with the metallicity of the host star. For giant planets, a dichotomy between dense and inflated planets is found. Denser planets ($R_p \lesssim$ 1.1 R$_{Jup}$) tend to be more massive as the metallicity of the host star increases, while inflated planets are more massive for less metallic hosts. If confirmed, this bimodality implies that the diversity of giant exoplanets depends on their Galactic birth locus, with dense giant planets being more numerous than inflated ones when \meta\ is higher than $\sim$1.5 times Solar, as in the central Milky Way regions.
}
   {The \Gaia\ spectroscopic catalogue of exoplanets and their host stars is large, homogeneous, and precise. Thus, it would be a useful added-value for planetary studies. Since it is based on literature data, it can also easily be updated thanks to future \Gaia\ data releases and other space- and ground-based surveys.}
   \keywords{Planetary systems - Stars: fundamental parameters, abundances – Galaxy: stellar content, disc - Planets and satellites: individual – Techniques: spectroscopic}
   
   \maketitle

\section{Introduction}
\label{Sec:Intro}

Exoplanetary science is experiencing rapid development thanks to the many surveys dedicated to the exoplanet detections and characterisations. With about 6000 exoplanets detected so far, it is becoming suitable to draw exoplanets populations and to link them with stellar properties. However, it is only with high-quality and homogeneous planetary parameters of  statistically significant samples that it is possible to understand the mechanisms leading to the formation and evolution of exoplanetary systems. In addition, such samples can be used to address the questions of exoplanets internal structure and their habitability. This kind of study moreover relies on accurate and precise parametrisation of the host stars. 
The two main exoplanets databases used so far, the Encyclopaedia of Exoplanetary Systems \footnote{http://exoplanet.eu/}(EES) and the NASA Exoplanet Archive\footnote{https://exoplanets.nasa.gov/discovery/exoplanet-catalog/} (NEA), provide a very rich set of exoplanetary and stellar properties; however, since they gather all data coming from diverse surveys and studies, they are not always consistent with each other. Thus, it is  complicated to use them directly to improve our understanding of exoplanet properties.

In this exoplanetary science context, the ESA \Gaia\ mission provides unique datasets of unprecedented high-quality and homogeneity.
Indeed, the last \Gaia\ release \cite[DR3][]{GaiaDR3} with its remarkable astrometric and photometric observations, was also accompanied by the first spectroscopic survey from space. This constitutes the largest-ever sample (by about one order of magnitude for the number of parametrised stars), even considering all ground-based facilities \citep{GSPspecDR3}. These authors  published a homogeneous catalogue of up to 5.6 million stars parametrised by their atmospheric parameters and chemical abundances thanks to the analysis of their \Gaia/Radial Velocity Spectrometer (RVS) spectra. This spectroscopic survey from space (called \gspspec,\ hereafter, from the name of the module in charge of this parameterisation within the \Gaia\ consortium)
has characterised, with high precision and a high number statistics, the Galactic stellar populations. It has been exploited for different purposes, among which Galactic Archaeology \cite[see e.g.][]{PVP_Ale}.  In this article, we present some of its possible contributions for the study of exoplanets. We note that part of the \gspspec\ catalogue
was already adopted by \cite{Swastik23} and \cite{Banerjee24} for studying the age distribution of about 800 Exoplanet Host Stars (EHSs) and the formation processes of exo-Jupiters, adopting  planetary parameters (e.g. masses) that could be inconsistent with the stellar ones (see below). We also note the article submitted by \cite{Berger23} that has a rather close goal as the present one. It however partly relies on \Gaia\ DR3 spectrophotometry metallicities that suffer from 'strong systematics errors', as pointed out by \cite{Berger23}. Thus, they have  to correct these metallicities and this hampers the precision of their derived stellar and planet parameters (and, thus, their conclusions). The present work should not be affected by these issues since the adopted parameters (in particular, the metallicity) mostly come from the \Gaia\ spectroscopic data that had  already been validated thanks to various analysis \citep[see e.g.][]{GSPspecDR3, Recio24} and it was not necessary to apply any ad hoc corrections.

The goal of this article is thus to present the largest catalogue of physical, chemical, and Galactic properties of EHSs that had been derived in a homogeneous way by \Gaia, complemented by some ground-based spectroscopic survey information. For instance, stellar luminosities associated with stellar radii and masses were directly computed from the collected 
stellar atmospheric parameters,  plus \Gaia\ photometry and parallaxes. 
Thus, they are  fully consistent between each other. In particular, we stress that stellar masses are derived without any use of stellar isochrones and are thus free from any stellar evolution model assumption. 
From these new stellar parameters, the exoplanetary parameters are rescaled carefully, according to the planet to star ratios taken from the literature. In other words, we do not modify or question the observations that led to the detection and characterisation of the exoplanets, but we used \Gaia\ derived EHS parameters to infer new exoplanetary ones. Our work is therefore fully independent and, thus, it is also complementary to other exoplanet catalogues that are much more homogeneous than the exoplanet databases, such as SWEET-cat \citep{SWEET1, SWEET2}; however, we note that the latter is based on a completely different spectra analysis
methodology and smaller statistics. Moreover, providing a large (and as  complete as possible) and homogeneous catalogue of EHS and planetary parameters allows us to carry out an improved study of their properties in a Galactic context, particularly within the framework of the \Gaia\ DR3 Galactic chemical cartography \citep{PVP_Ale}. Comparing homogeneous and large sample of planetary radii and masses also allows us to better explore their formation and evolutionary processes, in connection with their host star properties.

Our work is structured as follows. The EHS spectroscopic sample is built in Sect.~\ref{Sec:Samples} and the stellar atmospheric parameters, \AF\ content, luminosity, radius, and mass are derived in Sect.~\ref{Sec:Param}. 
The \Gaia\ spectroscopic catalogue of EHS is then presented 
and compared to the exoplanet online databases in Sect.~\ref{Sec:Catalog}. 
Next, Sect.~\ref{Sec:EHSpopulation} discusses the chemo-physical and Galactic properties of the EHS population.
In Sect.~\ref{Sec:Planets}, we derive the new \Gaia\ planetary properties and explore them in connection with their host star parameters. 
Our conclusions are summarised in Sect.~\ref{Sec:Conclu}.

\section{The EHS sample}
\label{Sec:Samples}

We built our EHS catalogue starting from the confirmed exoplanets listed in the EES and NEA (versions of July, 8, 2024). We merged these two lists, making sure not to count a system twice, but keeping confirmed exoplanets present in only one of the two databases. 
More details on this procedure are given in Sect.~\ref{sec:Gaia_cat_pl}.
We then discarded  objects for which neither the planetary radius, nor the planetary mass could be derived because of lack of observational data in these databases, along with the Solar System (which was in the database). We also rejected the exoplanets detected by micro-lensing since their host stars cannot be characterised. 
This resulted in a catalogue of 6,747 exoplanets\footnote{See Sect.~\ref{Sec:Planets} for the new determination of these exoplanet properties.}, orbiting around 4,840 different stars
(with several EHSs actually hosting multiple systems). We specify that this first sample contains 123 brown dwarfs that were in the NEA and EES databases and subject to the different filters.

To study these EHSs in terms of Galactic, stellar, and planetary physics, we first obtained their astrometric and spectroscopic properties.
For that purpose, we first looked for their \Gaia/DR3 identification (GDR3Id) and adopted them.
For most of them ($\sim$90\%), in part thanks to the SIMBAD database \citep{Simbad}, we were able to retrieve their GDR3Id from one of their most common stellar names provided by the above mentioned exoplanet databases. For the others, we preferred not to work with the coordinates provided by the exoplanet databases, since they were not accurate enough for cross-matching with the \Gaia\ catalogue.
Moreover, some of them are found to be high-proper motions star and/or multiple system  without any GDR3Id (or with a \Gaia\ DR2 Id not confirmed in DR3). These EHS were disregarded from the present study to avoid any contamination caused by a wrong identification.
Finally, we retrieved the GDR3Id of more than 4,300 
EHS, which constitutes our starting sample of stars. They are associated with about 5,400 
exoplanets.
We have checked that among them, the EHS discovered by \Gaia\ are included. These are the hot-Jupiter Gaia-1 (GDR3Id 3026325426682637824) and Gaia-2 (GDR3Id 1107980654748582144) detected by transit \citep{Panahi22}, and
HIP 66074 (GDR3Id 1712614124767394816) and
HIP 28193 (GDR3Id 2884087104955208064), the first \Gaia\ astrometric planet detections \citep{Arenou23}.\\

\begin{figure}[t]
    \centering
    \includegraphics[width=0.45\textwidth]{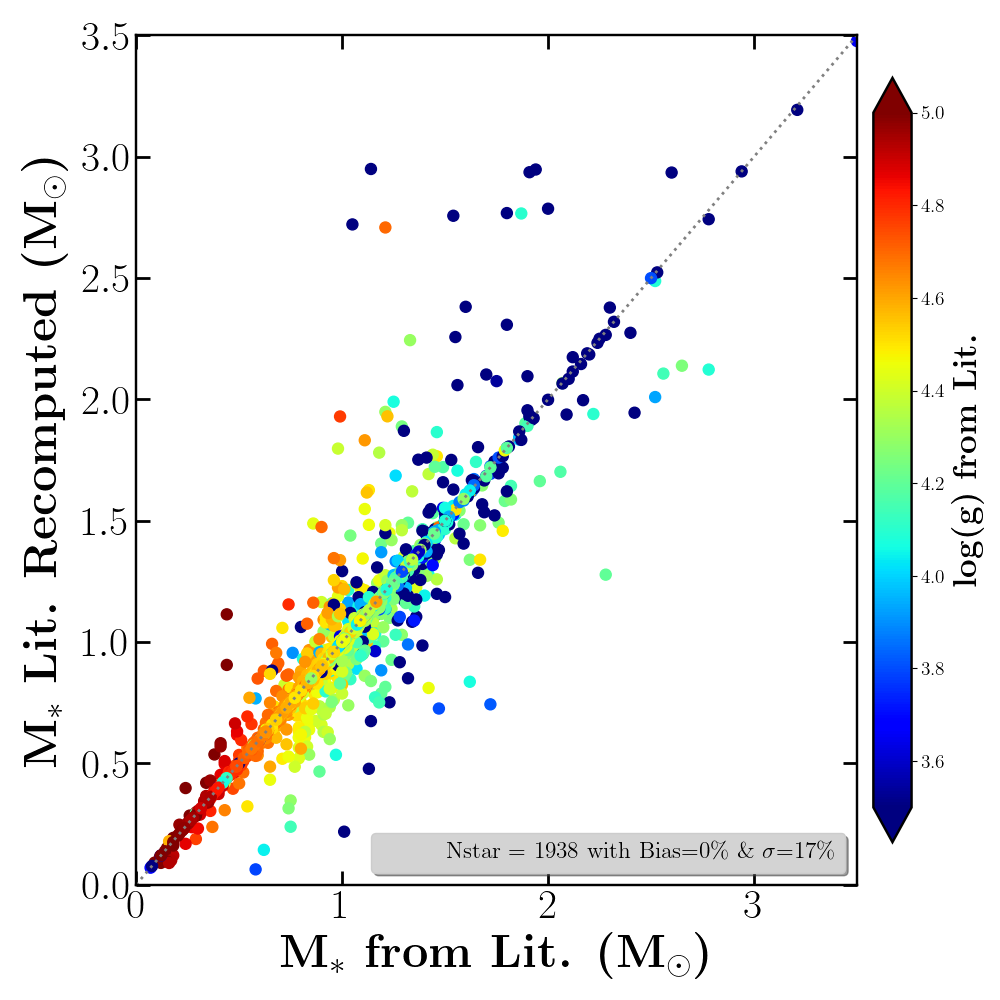} 
    \caption{Comparison between the database EHS masses with recomputed values adopting the database stellar radii and surface gravities (see text for details). The grey dotted line is the (1:1)-relation. If the database were fully consistent, all points would lie on this diagonal line. The most massive EHS are not shown for clarity reasons. 
    }
    \label{Fig:ParamLit}
\end{figure}

Finally, from this list of EHSs, it can be already shown that there can be inconsistencies among the stellar parameters found in the exoplanet databases. This is illustrated in Fig. \ref{Fig:ParamLit}, where we compare the database stellar masses  with recomputed values simply adopting the 
 database stellar radii and surface gravities (as described in Sect.~\ref{Sec:LMR}).
We could expect that the literature and recomputed values would be in perfect agreement if the online databases display a consistency among the parameters they contain. It can be seen that it is clearly not the case and we can state that there are severe issues when adopting indiscriminately the stellar parameters of the online databases. 
In particular,
the stellar surface gravities and radii can be strongly inconsistent with the stellar mass: more than 21\% and 8\% of the EHS have a recomputed mass that differ by more than 5\% and 20\%, respectively, with respect to the database values.
This clearly shows that a new determination of the stellar parameters (and subsequently the planetary ones) can be useful.

\section{Derivation of EHS parameters}
\label{Sec:Param}
In this section, we present the main properties of our EHS sample derived from
the \Gaia/\gspspec\ module and complemented by ground-based spectroscopic surveys for those not found in \gspspec. All these data have already been extensively validated and we refer to the corresponding literature provided below for that purpose. In brief, we infer the main stellar atmospheric parameters (\T, \g, \meta, and \AF) from the analysed spectra (Sect.~\ref{SubsectTGM} \& \ref{SubsectAFe}) and then the stellar luminosities, radii, and masses are derived using additional photometric and astrometric data (Sect.~\ref{Sec:LMR}).  

\subsection{Stellar atmospheric parameters: \T, \g,\ and \meta}
\label{SubsectTGM}
First, among all the 
identified EHS, 
$\sim$42\% were fully parametrised (i.e. their \T, \g, \meta,\ and chemical abundances have been provided) by the \gspspec\ module thanks to the analysis of their \Gaia/RVS spectra. We adopted these effective temperatures plus their calibrated (see below) surface gravities and global metallicities\footnote{
The \gspspec\ metallicity parameter \meta\ traces the [Fe/H] abundance \citep{GSPspecDR3}.} 
in the present study.
We refer to Sect.~\ref{Sec:Catalog} for a discussion on the \gspspec\ quality flags (QF), which have to be used to select the best parametrised stars.

Then, we searched if the other EHS have been parametrised by ground-based spectroscopic surveys. Another 790 (18\%), 40 (1\%), and 7
stars were actually found in the  APOGEE-DR17 \citep{APOGEEDR17}, GALAH-DR3 \citep{GALAHDR3}, and \Gaia-ESO \citep[GES,][]{GESGilmore, GESRandich} surveys, respectively, and not in \gspspec. These complementary surveys provide additional stellar parameters and chemical abundances derived from higher resolution and broader spectra than \gspspec, but for a statistics that is much smaller (one order of magnitude, typically).  We note that some stars were found in more than one survey and that all the parameters are not available for some of them. For such cases, we adopted in priority order their parameters provided first by \gspspec, and then by APOGEE, GALAH, and GES to build a catalogue as homogeneous as possible, since
\gspspec\  is the catalogue with the most EHS, followed by APOGEE.
We added these ground-based characterised EHS to build our final working sample of
EHS having a spectroscopic parameterisation\footnote{We warn that all the parameters are not available for all the stars (see Sect.~\ref{Sec:Catalog}).} (60\% of the whole sample with GDR3Id). 
For the APOGEE stars, we also adopted their calibrated stellar parameters. All these EHS atmospheric parameters were not derived with the same quality, depending on the stellar types, the spectra quality and signal-to-noise ratio (\SNR), and the survey characteristics, among others.  We therefore considered the most appropriate QF associated with each ground-based survey. In practice, the APOGEE parameterisation quality was examined thanks to its {\it PARAMFLAG} and {\it ASPCAPFLAG} quality flags. For GALAH, we looked at their {\it flag$\_$sp}, {\it flag$\_fe\_h$} and {\it flag$\_Ca\_fe$} flags. Finally, the {\it Simplified Flag} was considered for the GES star (see Sect.~\ref{Sec:Catalog} for an explanation of these QF use within the present project).

Since most of these EHS have spectroscopic data provided by the \gspspec\ catalogue and cover a large range in \T, we have homogenised their atmospheric parameters to put all of them on the \gspspec\ scale.
For that purpose, the \gspspec\ \g\ and \meta\ were calibrated as a function of \T\ thanks to the relations recommended by \citet[][see their Table~A.1]{Recio24}. In addition, to further increase the accuracy in \g , a second order calibration considering gravity and metallicity effects was applied following de Laverny et al. (in prep.).
We recall that these calibrations were derived during the validation step of the \gspspec\ catalogue thanks to a comparison with ground-based spectroscopic surveys \citep[APOGEE and GALAH in particular, see][]{GSPspecDR3}. These calibrations were found to strongly reduce the possible biases between the \Gaia\ spectroscopic survey and the ground-based ones. No calibration was found to be necessary for \T\ in the study by \cite{GSPspecDR3}. 

Among the 1,812 EHS parameterised by \gspspec, about 350 
have also APOGEE parameters\footnote{We recall that we adopted in the following analysis the \gspspec\ data of these common stars.}. This allows us to check the homogeneity.
For the 84 EHS 
in common 
with a spectrum \SNR$_{RVS}$>100, we found a median and median absolute deviation of (+20, 77~K), (-0.04, 0.13~dex), and (-0.01, 0.06~dex) in \T, \g,\ and \meta, respectively, in the sense \gspspec-APOGEE.
Such biases
are rather negligible and we therefore did not apply any additional calibration.
The number of EHS in common with GALAH and GES is too small to conduct any comparison.
We refer to \cite{GSPspecDR3} and \cite{Mathieu24} for a confirmation that both surveys agree well with \gspspec.

\subsection{The \AF\ chemical index}
\label{SubsectAFe}
We retrieved the \alfa-element abundance relative to iron (\AF)\footnote{In addition to help characterising the EHS chemical content, this index is also required to compute the stellar luminosities (see below).} for any \gspspec\ EHS. As for the atmospheric parameters, \AF\ were calibrated with the third-order polynomial as a function of \T\ derived by \cite{GSPspecDR3}\footnote{See~also~\url{https://www.cosmos.esa.int/web/gaia/dr3-gspspec-metallicity-logg-calibration}}. Such a calibration was estimated by assuming that the abundance distributions should be close to zero in the Solar neighbourhood for stars with metallicities close
to Solar and velocities close to the Local Standard of Rest \citep[LSR, see the third panel of Fig.14 of][confirming the good agreement between the calibrated \gspspec\ \AF\ and APOGEE \CaFe]{GSPspecDR3}. Finally, as recommended by \cite{GSPspecDR3}, we filtered out the \AF\ value of two hot metal-poor supergiants defined by \T>4000~K and \g<1.5 since their chemical index is found outside the \gspspec\ reference grid.

For the EHS with ground-based parameters, we adopted their \CaFe\ as a proxy of \AF\ since, in the \gspspec\ analysis of the RVS spectra, the main \alfa-element spectral signatures come from the calcium IR triplet lines.
The median and MAD of the EHS \AF\ found both in \gspspec\ and APOGEE and with \SNR$_{RVS}$>100 are very small, confirming the excellent agreement.

\subsection{Stellar luminosities, radii, and masses}
\label{Sec:LMR}

From the above described spectroscopic parameters that are fully consistent within each others and in very good agreement with ground-based large spectroscopic surveys, we computed the luminosity ($L_\star^\Gaia$) and spectroscopic mass ($M_\star^\Gaia$) and radius ($R_\star^\Gaia$) for each star
thanks to the procedure detailed in de Laverny (in prep.)\footnote{In the following and to ease the comparison with literature values, all these parameters are noted with a superscript "\Gaia".}.
It is important to note that the above quantities are independent of any stellar evolutionary models and Galactic priors.

Shortly, we first estimated 
the absorption in the $G$-band ($A_G$) by comparing the \Gaia\
($B_p-R_p$) observed colours to theoretical ones estimated thanks to the \cite{Luca21} relation between stellar colour and atmospheric parameters. For that purpose, we adopted the spectroscopic \T, \g , \meta,\ and \AF\ values defined in the above sub-sections. Then, the absolute magnitudes in the \Gaia\ $G$-band were computed from the \Gaia\ DR3 $G$-magnitudes and the \cite{Coryn21} geometric distances. The absolute bolometric magnitudes are then easily obtained, adopting the bolometric corrections of \cite{2018MNRAS.479L.102C}, which were also computed from the spectroscopic atmospheric parameters.
Moreover, the ratio between stellar and Solar luminosities is computed by adopting the Solar absolute bolometric magnitude $M_{\rm bol}^\odot$=4.74 \cite[IAU 2015 resolution B2,][]{IAU}. These stellar luminosities are thus fully consistent with the spectroscopic atmospheric parameters. Their accuracy is mostly related to the \Gaia/DR3 parallaxes and we refer to de Laverny et al. (in prep.) for more details about their derivation and validation.
Finally, the stellar radius is directly obtained from  $L_\star^\Gaia$ and \T\  via the well-known relation: 
\begin{equation}
   L_\star^\Gaia = 4\pi (R_\star^\Gaia)^2\sigma T_{\rm eff}^4,
   \label{eq:Lstar}
\end{equation}
with $\sigma$ being the Stefan-Boltzmann constant.
Similarly, the stellar mass is derived from the surface gravity:
\begin{equation}
   M_\star^\Gaia = (R_\star^\Gaia)^2 \times (g/g_\odot),
   \label{eq:Mstar}
\end{equation} where $M_\star^\Gaia$ and $R_\star^\Gaia$ are expressed in Solar units and, \T$_{,\odot}$=5777~K, and log($g_\odot)$=4.44.
For all these quantities, the uncertainties were estimated by performing 1,000 Monte-Carlo
realisations, propagating the uncertainties on each atmospheric parameter (that mainly reflect the spectral \SNR), distance
and \Gaia\ magnitudes in the different bands. 
The published values are the median of the distributions and their associated uncertainties are half of the difference between the 84$^{th}$ and 16$^{th}$ quantiles
of the distributions, hence corresponding to a 1-$\sigma$ uncertainty for a normal distribution.
The quality of the $L_\star^\Gaia$,  $M_\star^\Gaia$, and  $R_\star^\Gaia$ determinations can be assessed thanks to a quality flag ($Flag_{\rm Abs}$), presented in Sect.~\ref{Sec:Catalog}.

We refer to de Laverny et al. (in prep.) and \cite{Recio24} for an extensive validation of the computed stellar radii and masses \citep[see also][]{PatGamDor}. These studies compared their computed and literature values derived either from interferometric or asteroseismic data. The agreement was found to be excellent (almost no biases) and typical dispersions are smaller than 5\% and 15\% for radii and masses, respectively.  Moreover, these dispersions are strongly reduced when \T\ and \g\ are in good agreement between the compared studies. Finally, another specific radius and mass validation for EHS with asteroseismic information can be found in Appendix~\ref{Appendix:Valid}.

In summary, the above comparisons confirm the high-quality of our spectroscopic stellar masses and radii, particularly when our quality flags are adopted for selecting the highest-quality data and when the agreement between the adopted effective temperatures and surface gravities is good. We also emphasise that filtering with our uncertainties could help to build a EHS sub-sample with high-precision values of  $M_\star^\Gaia$ and  $R_\star^\Gaia$. 

\subsection{Galactic kinematic and dynamical properties}
Lastly, we  computed the Galactic positions and kinematic plus dynamical properties of the EHS. 
They were estimated following a similar procedure as in \cite{PVP_Ale}.
We first computed their Galactic positions, thanks to \Gaia\ coordinates and geometric distances of \cite{Coryn21}. Distances were not available for $\sim$9\% of the sample stars for which no Galactic properties were therefore published\footnote{Naturally, no luminosity, radius and mass were also inferred for these stars.}.
The kinematic properties were then estimated by adopting the \Gaia~DR3 radial velocities \citep[\Vrad,][]{David23} and \Gaia~eDR3 proper motions. 
The LSR velocity at the Sun's position was set at  $V_{\rm LSR}$=238.5~km.s$^{-1}$ \citep{SchBinDeh2010}, while a distance to the Galactic centre of $R_0=8.249$~kpc \citep{Gravity20} has been adopted.
Finally, the Galactic orbital parameters 
were computed as in \cite{Pedro23} and are fully compatible with the other Galactic data. We note that 37 EHS do not have any published \Gaia/DR3 \Vrad\ and, hence, no kinematic or dynamical properties. For homogeneity, we preferred not to complete this sample by looking for other possible literature radial velocities for these stars.

\section{The \Gaia\ spectroscopic catalogue of exoplanet host stars}
\label{Sec:Catalog}

\begin{table}[t]
        \caption{\label{Tab:EHS} \Gaia\ spectroscopic catalogue of EHS.}
        \centering
        \begin{tabular}{ll}
        \hline
        Label & Description \\
        \hline
        GDR3Id & $Gaia$ DR3 Identification\\
        StarName &  Stellar name in the EHS databases\\
        N\_Exop & Number of detected exoplanets \\
        SpecSurvey & Spectroscopic surveys: 1=\gspspec \\
        & 2=APOGEE, 3=GALAH, 4=GES\\
        Teff & EHS effective temperature (K)\\
        Teff\_err &  \T$^\Gaia$\ uncertainty (K) \\        
        logg &  EHS surface gravity in log ($g$ in cm.s$^{-2}$) \\
        logg\_err & \g\ uncertainty\\        
        Meta & EHS mean metallicity (\meta$\sim$[Fe/H],  in dex)\\
        Meta\_err & \meta\ uncertainty (dex) \\
        AFe & EHS enrichment in \AF\ ($\sim$[Ca/Fe], in dex)\\
        AFe\_err & \AF\ uncertainty (dex) \\
        A\_G & Absorption in the $G$-band (mag.) \\
        A\_G\_err & A$_G$ uncertainty (mag.) \\
        L & EHS luminosity $L_\star^\Gaia$ ($L_\odot$)\\
        L\_err &  $L_\star^\Gaia$ uncertainty ($L_\odot$) \\
        R &  EHS radius $R_\star^\Gaia$ ($R_\odot$) \\
        R\_err &  $R_\star^\Gaia$  uncertainty ($R_\odot$)  \\
        M &  EHS mass $M_\star^\Gaia$ ($M_\odot$) \\
        M\_err &   $M_\star^\Gaia$ uncertainty ($M_\odot$) \\
        X &  $X$ Galactocentric Cartesian coordinate (kpc) \\
        Y &  $Y$ Galactocentric Cartesian coordinate (kpc) \\
        Z &  $Z$ Galactocentric Cartesian coordinate (kpc) \\
        V\_R &  Galactocentric radial velocity (km.s$^{-1}$) \\
        V\_phi &  Azimuthal Galactic velocity (km.s$^{-1}$) \\
        V\_Z &  Galactocentric vertical velocity (km.s$^{-1}$) \\
        Z\_max &  Max. abs. distance to the Galactic plane (kpc) \\
        R\_peri &  Orbital pericentre radius (kpc) \\
        R\_apo  &  Orbital apocentre radius (kpc) \\
        ecc &  Orbital eccentricity\\
        Flag\_Param &  QF associated with the EHS parameterisation \\
                    &  (0/1/2) = (High/Good/Low)-Quality\\
                    & (9) No \T\ available \\
        Flag\_Abs &  QF associated with absorption estimate\\
                  &  (0/1) = (high/good)-qual. (9)=LMR filtered\\
         \hline
        \end{tabular}
        \tablefoot{The full version of this table is available in electronic form at the CDS.}
\end{table}

\subsection{Content of the catalogue}
All the computed stellar properties are provided in an electronic table whose content is detailed in Table~\ref{Tab:EHS} (see Sect.~\ref{Sec:Planets} and Table~\ref{Tab:ExoP} for a presentation of the derived planetary properties).
We recall that one of the main values of this catalogue is its extremely high homogeneity for a statistically significant large sample of EHSs. This homogeneity allows us to avoid systematics and biases that could be hidden when combining data from different observations and analysis strategies. 

In particular, this homogeneity is ensured by the direct calculations presented above, which do not involve stellar evolution models. Different stellar evolution models can yield different stellar fundamental parameters for the same observables and this external source of error is often not taken into account. Hence, the uncertainties provided in this catalogue may also be more realistic. Besides, as we  show below, the relative uncertainties on the stellar radii and masses are generally smaller than found in the literature.

In the construction of the final catalogue presented in Table~\ref{Tab:EHS}, particular attention was paid to the quality of the stellar parameters and some too low-quality data were filtered out or flagged. For instance, the luminosities
are of great importance since several other stellar and planetary parameters are directly derived from  $L_\star^\Gaia$. 
Low-quality  $L_\star^\Gaia$ could indeed prevent a clear interpretation of the planetary properties. For this reason, we did not publish $L_\star^\Gaia$ and related stellar and Galactic properties for
242 stars (170, 68, 3, and 1
from \gspspec, APOGEE, GALAH, and GES, respectively),  with  astrometric $ruwe$ parameters larger than 1.4 -- indicating a suspicious solution and, consequently, a possible inaccurate estimated distance.
Moreover, we did not publish  $L_\star^\Gaia$,  $R_\star^\Gaia$, and  $M_\star^\Gaia$
for 15 (3, 11, and 1 from \gspspec, APOGEE, and GALAH, respectively)
other stars whose relative uncertainty in the luminosity was too high (>40\%). 
On the other hand, the surface gravity of some stars was found to be less precise, particularly when their values are close to the borders of the reference grid adopted during the \gspspec\ parameterisation. 
As a consequence, we did not publish their \g\ and, hence,  $M_\star^\Gaia$ since it varies linearly with $g$ as indicated in Eq.~\ref{eq:Mstar}. Moreover, the mass of stars having a \g\ uncertainty larger than 0.3~dex was also not published since this corresponds to a relative mass error larger than a factor 2 (i.e. $\sim$100\% for $M_\star^\Gaia$, see Eq.~\ref{eq:Mstar}).  
Finally, all the stellar parameters of 24 cool-giant stars with a \gspspec\ $KM$-quality flag larger than zero (indicating issues in their parametrisation) have only their Galactic properties published.

In summary, Table~\ref{Tab:EHS} contains in total 2,534 
EHSs ($\sim$87\% being dwarfs) and,
among them, 
1,812, 731, 25, and 5
are (at least partly) parametrised by \gspspec, APOGEE, GALAH, and GES spectroscopic surveys, respectively.  \T\ and $L_\star^\Gaia$ (and, hence,  $R_\star^\Gaia$) are published for 98\% and 80\% of them whereas  $M_\star^\Gaia$ is available for 69\% of the whole sample. We also provide the Galactic positions and kinematic and orbital information for 91\% and 90\% of the whole sample, respectively, given that their distances and radial velocities are available.\\

Finally, in order to assist future users in the selection of stars having the most optimal quality parameters, two quality flags (QFs) related to the stellar properties  defined and added in Table~\ref{Tab:EHS}:
\begin{enumerate}
    \item $Flag_{\rm Param}$ summarises the information provided by the different spectroscopic surveys about their parameter determination quality, thanks to their respective QF. Its possible values are '0', '1', and '2' for high-, good-, and low-quality parameterisation, respectively, while $Flag_{\rm Param}$=9 for the very few stars without available effective temperature.
Firstly, we considered the 13 first \gspspec\ QF that refer to the three main atmospheric parameters \citep[see Sect. 8 of][for more details]{GSPspecDR3}. For the \gspspec\ EHSs, $Flag_{\rm Param}$ is set to 0 when all the \gspspec\ QF are null. It is equal to 1 or 2 when at least one of the QF is equal to or larger than 1, respectively.
    Most stars with $Flag_{\rm Param}$=2 are the results of a low quality value of their \gspspec\ $extrapol$ QF\footnote{This $extrapol$ flag refers to the extrapolation level of the parametrisation \cite[see][for more details]{GSPspecDR3}.}. Secondly, APOGEE EHSs were examined thanks to their $PARAMFLAG$ and $ASPCAPFLAG$ QF. If no warning are associated to these flags, then $Flag_{\rm Param}$=0.
    If a warning is set for one of them, then $Flag_{\rm Param}$=1 and $Flag_{\rm Param}$=2 if a low-quality determination is identified. Thirdly, when all the GALAH {\it flag$\_$sp}, {\it flag$\_fe\_h$} and {\it flag$\_Ca\_fe$} flags are null, we fixed $Flag_{\rm Param}$=0. Only three stars are identified with a low-quality parameterisation and have been set to $Flag_{\rm Param}$=2.
Finally, the few EHSs parameterised by GES have a medium quality {\it Simplified Flag} and, hence, their $Flag_{\rm Param}$=1. In practice, the 1,557 
    EHSs with $Flag_{\rm Param}=0$ are associated with very high quality parameters but
    at the expense of a lower statistics. This QF can, however, be safely relaxed up to $Flag_{\rm Param}<2$ to increase the sample size up to 2,024    stars.      
    \item $Flag_{\rm Abs}$ 
    assesses the quality of the  $L_\star^\Gaia$,  $R_\star^\Gaia$ and  $M_\star^\Gaia$ calculations. Its three possible values are '0', '1', and '9' for the high-, good-, and bad-quality determinations, respectively.
    $Flag_{\rm Abs}$ = 9 most often results from bad extinction and/or unreliable bolometric correction estimates.
    We therefore filtered out in Table~\ref{Tab:EHS} the stellar luminosity, radius, and mass (and, hence, the planetary radius and mass since they are derived from the stellar values) of the 333 
    EHSs with a $Flag_{\rm Abs}$ equal to 9. Practically, 2,240 
    stars with $Flag_{\rm Abs}<9$ can be safely selected.
\end{enumerate}

Finally, for the following discussions and in line with the recommended use of these quality flags, we have defined a high-quality sub-sample of EHSs (noted $HQ$, hereafter) with $Flag_{\rm Param}<2$ and $Flag_{\rm Abs}<9$. This $HQ$ sample contains 1,826 
EHSs (71 \% of the whole sample), 
$\sim$84\% of them coming from \gspspec. We also recommend again to additionally consider the reported uncertainties associated with the different parameters to define other high-quality sub-samples (see e.g. Fig.~\ref{Fig:MetaAlpha}).

\subsection{Comparison between the newly derived EHS properties and the exoplanet databases}
\label{sub:Compar}

For our EHSs, we now have a new set of their physical properties estimated thanks to homogeneous spectroscopic and \Gaia\ DR3 data, as described above. It is of interest to compare these new values with those previously published in the literature and that can be retrieved in the online EES and NEA tables (noted hereafter with the superscript 'Lit.'). 
These literature values are indeed based on different methodologies. For instance, they are most often based on isochrones and stellar evolution models, 
whereas our approach is fully independent of such models.
Moreover, a different stellar radius or mass directly affects the planet ones.  It is up to the user to choose stellar parameters that best suits their needs but we need to be aware of the external errors, which are linked to different methodologies that  may give different results, each with an associated small internal error \citep[see e.g.][]{Crida-etal-2018}. Comparing our \Gaia\ spectroscopic catalogue of EHSs with those available online allows us to illustrate this sensitivity. In addition, we discuss the compatibility and respective relevance of these two datasets.

\subsubsection{Stellar radius, luminosity, and effective temperature}

\begin{figure}[t]
    \centering
    \includegraphics[width=0.5\textwidth]{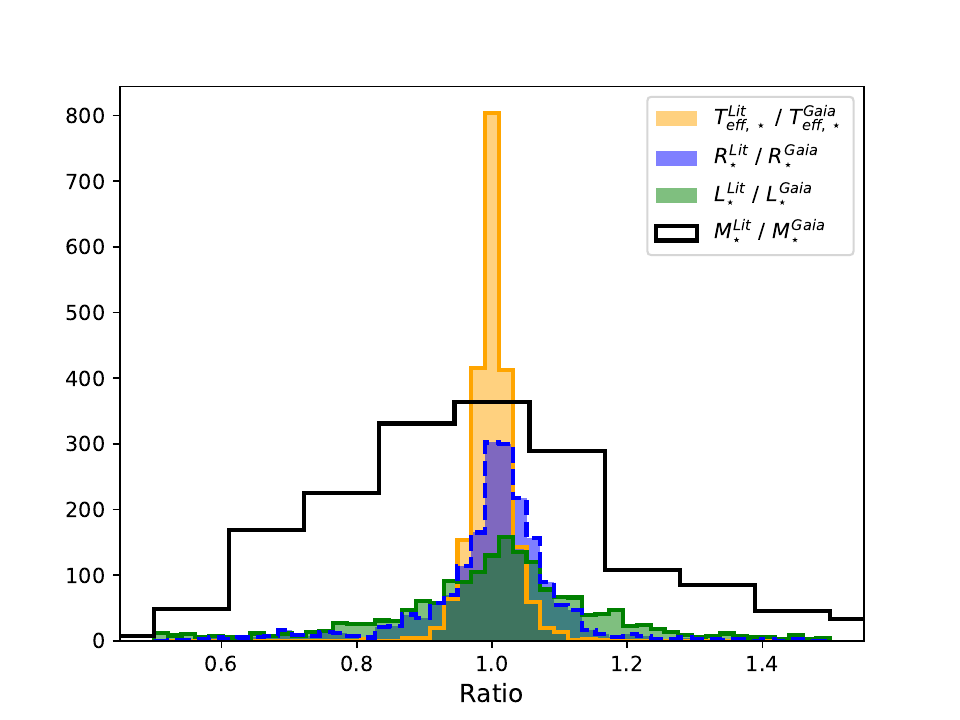} 
    \caption{Distributions of the parameters ratio. Blue: $R_\star^{Lit.}/R_\star^\Gaia$. Orange: $\Tm^{Lit.}/\Tm^\Gaia$. Green: $L_\star^{Lit.}/L_\star^\Gaia$. Black: $M_*^{Lit.}/M_\star^\Gaia$.}
    \label{Fig:Ratio}
\end{figure}

These three quantities are connected thanks to Eq.~\ref{eq:Lstar}.
As explained above, $L_\star^\Gaia$ and \T$^\Gaia$ are computed rather independently, and $R_\star^\Gaia$ is then derived from them. Conversely, we have taken $R_\star^\Lit$ and \T$^\Lit$ from the online tables, and calculated $L_\star^\Lit$ using Eq.~(\ref{eq:Lstar}) since it is not present in the literature\ tables. Figure \ref{Fig:Ratio} shows the distribution of the ratios of these quantities found in the literature and in the complete \Gaia\ EHS catalogue. The agreement is good for the radius, temperature, and luminosity. 
Below, we analyse these aspects in more detail.

\paragraph{Stellar radius}
\begin{figure}[t]
    \centering
    \includegraphics[width=0.5\textwidth]{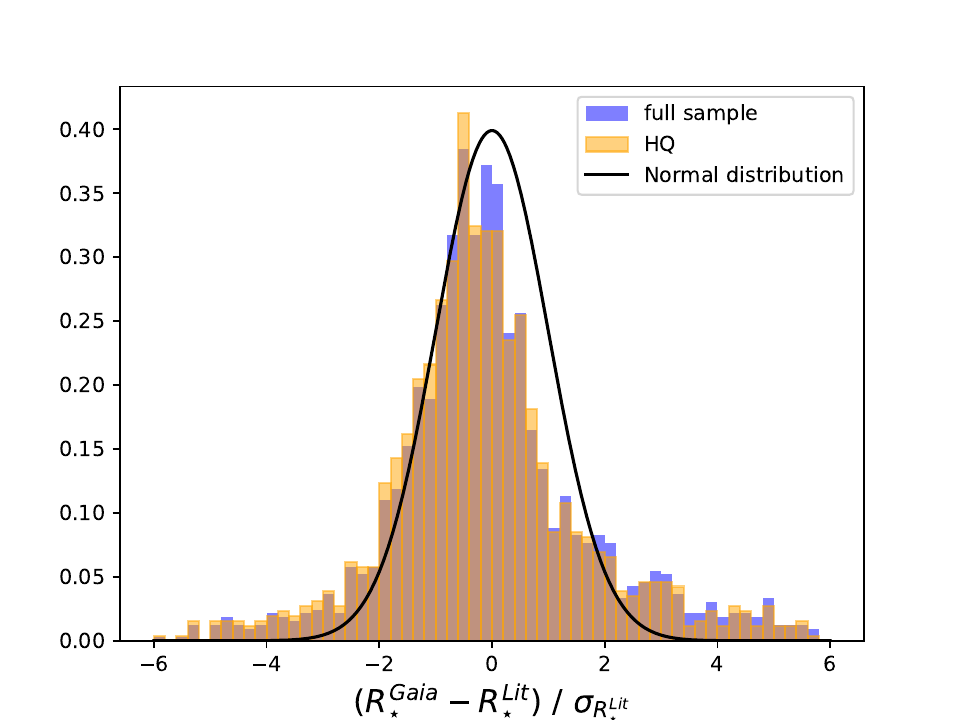} 
    \caption{Distribution of $(R_{\star}^\Gaia-R_\star^{\Lit})/ \sigma_{R_\star^\Lit}$.}
    \label{Fig:R_norm}
\end{figure}
The ratios between our stellar radii and those of the online exoplanet tables have a median of 0.992 (average of 1.028) with a standard deviation of 0.178 for 1715 
stars. 
If we restrict our sample to the $HQ$ stars in our catalogue, the median ratio becomes 0.990 (average 1.017) and the standard deviation 0.142, for 1374 
stars.
More than 80\% $HQ$ stars have \RstarGaia\ within $\pm10\%$ of \RstarLitt.
Considering the covered stellar radii range from $\sim$0.3 to $\sim$100~$R_\odot$, this is a very good agreement. However, we recall that a difference of $15\%$ in a planetary radius has strong consequences on the inferred planetary properties ($50\%$ in the volume for instance).

Moreover, how does this difference compare with the uncertainty? In Fig.~\ref{Fig:R_norm}, we show the histogram of $(R_\star^\Gaia-R_\star^{\Lit})/\sigma_{R_\star^\Lit}$. If the uncertainty $\sigma_{R_\star}^{\Lit}$ was correctly estimated and if $R_\star^\Gaia$ was the true value of the stellar radius, then this histogram should be a normal distribution (Gaussian with mean $0$ and standard deviation $1$), by definition. As shown in the figure, it is not far from this. Actually, the standard deviation of this quantity is 2.5, but it is pulled to a large value by a few points; excluding the points beyond $\pm 3$, the standard deviation decreases to 1.18. In other words, the stellar radii provided by our \Gaia\ spectroscopic EHS catalogue are very much compatible with the online table,
although the adopted methodology can differ strongly.

Conversely, if now we plot $(R_\star^{\Lit}-R_\star^\Gaia)/\sigma_{R_\star}^\Gaia$ (i.e. we swap the roles of the two datasets), we find a much wider distribution. This can be explained by the fact that the uncertainty in the \Gaia\ catalogue is roughly 3 times smaller than in the online tables. The median of the ratio $\sigma_{R_\star}^\Gaia/\sigma_{R_\star}^{\Lit}$ is $0.35$. As a consequence, the values of the literature stellar radii (without considering their uncertainty) appear incompatible with our Gaia estimates.

Although it is not possible to obtain the true stellar radii at present, it seems that our catalogue provides stellar radii with a higher level of precision than usual and presumably accurate. 
The relative error $\sigma_{R_\star}^\Gaia/R_\star^\Gaia$ follows an exponential probability density function of parameter $\sim$0.027, versus $\sim$0.075 for the literature. Thus,  our stellar radii might be useful in looking for planetary radii with a $3\%$ uncertainty -- in cases where the transit light curve has been measured accurately.

\begin{figure}[t]
    \centering
    \includegraphics[width=0.5\textwidth]{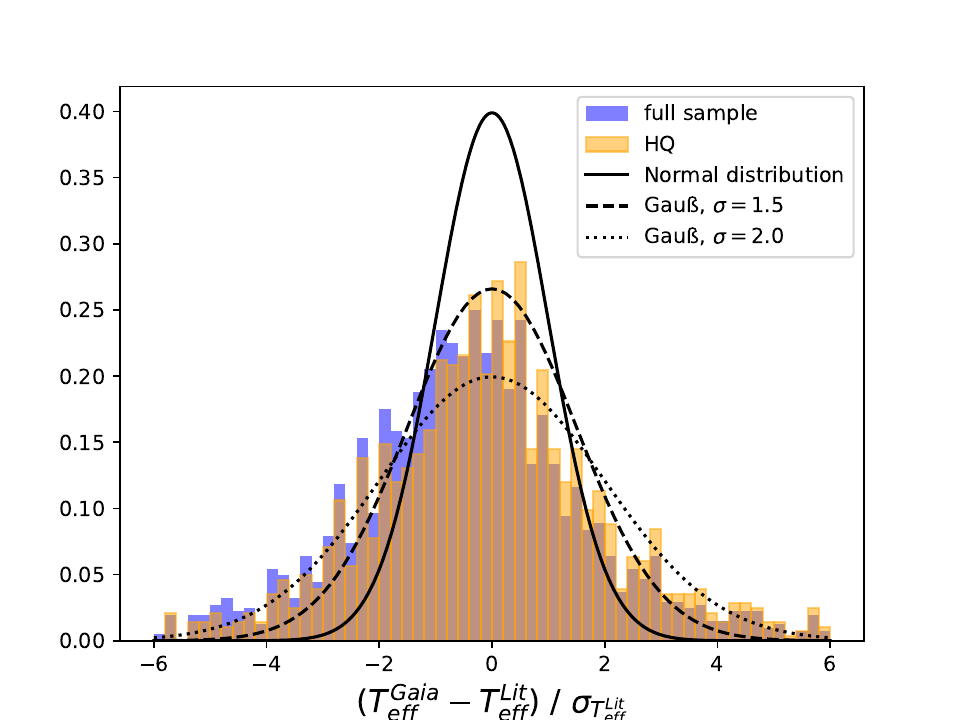} 
    \caption{Distribution of $(\Tm^\Gaia-\Tm^{\Lit})/\sigma_{\Tm}^{\Lit}$.}
    \label{Fig:T_norm}
\end{figure}

\paragraph{Effective temperature}
Concerning the stellar temperature, the ratio of the values in our catalogue and in the online tables displays a median of 0.995 and a mean of 0.994 for 2286 EHSs which have a \T\ in both datasets, with a standard deviation of 0.039 (see Fig. \ref{Fig:Ratio}). For the 1631 $HQ$ stars present in both catalogues, we get a mean of 1.001, median 0.999, and standard deviation 0.032. 
More than 99\% $HQ$ stars have the \Gaia\ and literature\ temperatures within 10\%.
This means that our catalogue and the online tables are in excellent agreement for \T. But similarly as for the radius, we find that the \T\ uncertainty is smaller for our catalogue, which therefore helps gaining precision. The median ratio between the uncertainties in our \Gaia\ spectroscopic EHS catalogue and those in the literature is $0.35$. 
In summary, our catalogue seems  to be more precise and not less accurate than the literature values. Again, we can plot the histogram of $(\Tm^\Gaia-\Tm^{\Lit})/\sigma_{\Tm}^{\Lit}$, as seen in Fig. \ref{Fig:T_norm}. The agreement with the normal distribution is less good (standard deviation: 4.1 for the whole sample, 1.4 for the $HQ$ cases at less than $3\sigma$), showing that the uncertainties are probably underestimated.

\begin{figure}[t]
    \centering
    \includegraphics[width=0.5\textwidth]{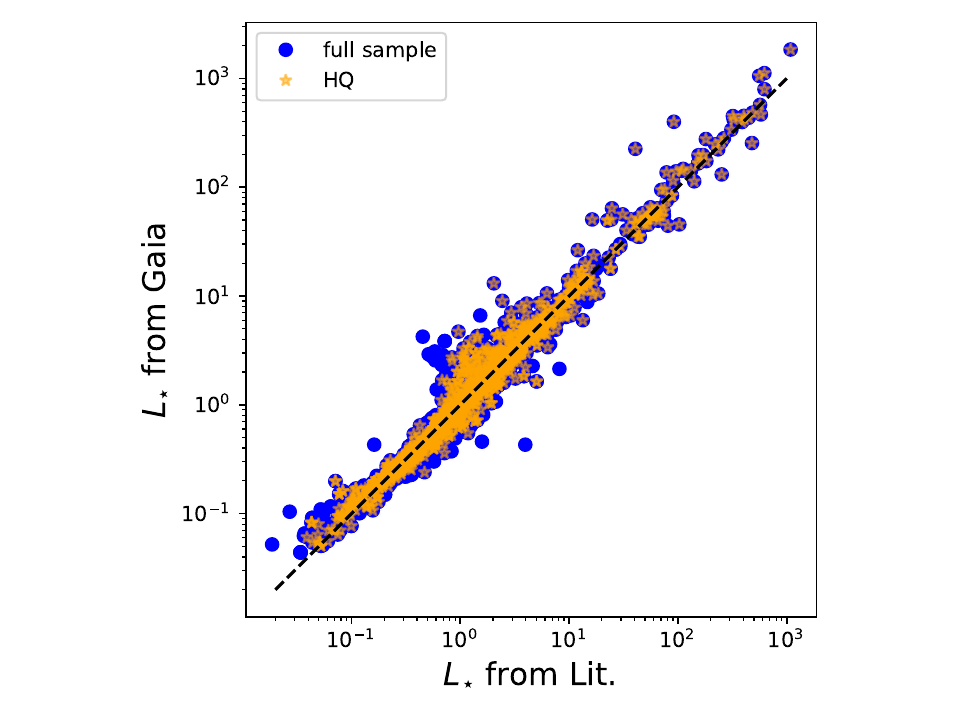} 
    \caption{Stellar luminosity comparison between our \Gaia\ spectroscopic catalogue and the data available in the online exoplanet tables. Blue dots correspond to the whole sample while orange dots are the stars with high quality data.}
    \label{Fig:LL}
\end{figure}

\paragraph{Luminosity}
For 1698 stars having $L_\star$ in both catalogues, the median of the ratio is $0.981$ (mean 1.079), but with a standard deviation of 0.513. For the 1362 ones with the $HQ$ data, the median becomes 0.984 and the standard deviation decreases to 0.381, with 56\% stars within 10\%. The larger standard deviation than in previous cases should be expected from Eq.~(\ref{eq:Lstar}): with the radius squared and the effective temperature to the power 4, the uncertainties are magnified. 
Nonetheless, the agreement is rather good, for a parameter that spans more than four orders of magnitude, as shown in Fig.~\ref{Fig:LL}. 

\subsubsection{Stellar mass}

\begin{figure}
    \centering
    \includegraphics[width=0.5\textwidth]{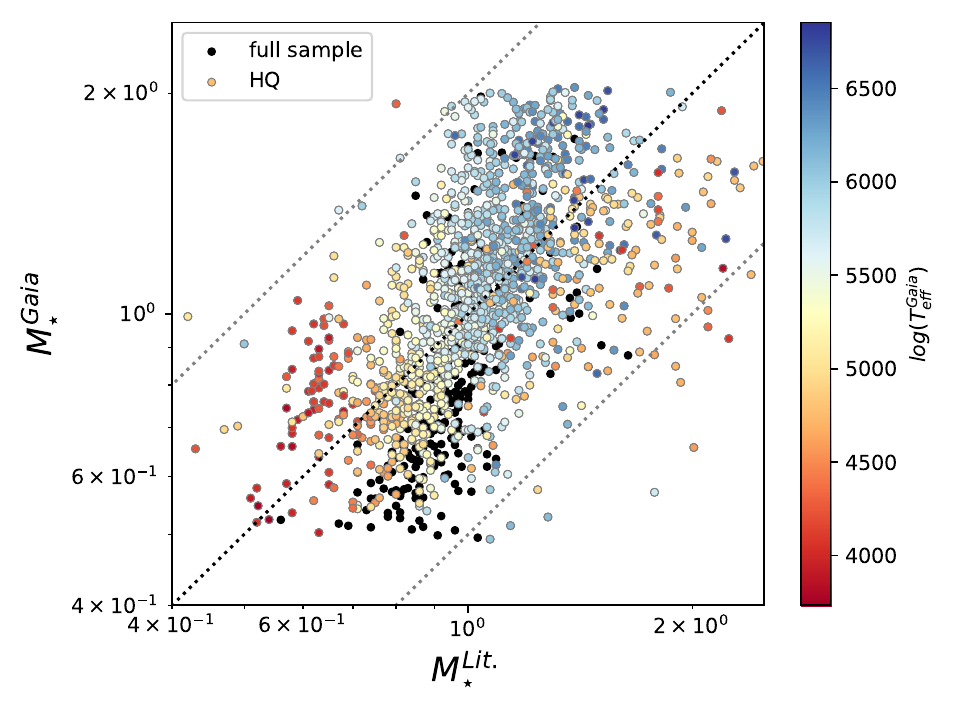} 
    \caption{Stellar mass estimated from our Gaia catalogue ($y$-axis) versus the one given in the exoplanet online tables ($x$-axis). The black dots correspond to the full sample, while the $HQ$ sample is colour-coded by $T_{\rm eff}^\Gaia$.
    The grey dotted lines show ratios of $2$ and $1/2$ as visual reference. We refer to Fig.~\ref{Fig:Ratio} to see the distribution of the $M_\star^\Gaia/M_\star^\Lit$ ratio.
    }
    \label{Fig:M}
\end{figure}

As for the stellar mass, the agreement is much less tight, as shown by the black histogram in Fig.~\ref{Fig:Ratio}. For the 1658
stars found in both catalogues, the mass ratio has a median of 
1.001, a mean of 1.034, and a standard deviation of 0.283.
If we restrict our sample to the 1382 stars with $HQ$ parameters, the mean becomes 1.063, the median 1.024, and the standard deviation 0.284.
Only less than 37\% have \MstarGaia\ within 10\% of \MstarLitt.
No major bias is noticed, but the dispersion is quite large, especially considering that most stars are within a factor two of the solar mass.
The scatter plot shown in Fig.~\ref{Fig:M} of the stellar mass in both catalogues illustrates such a disagreement, and highlights that the spread is larger in the \Gaia\ catalogue than in the literature.

\begin{figure}
    \centering
    \includegraphics[width=0.48\linewidth]{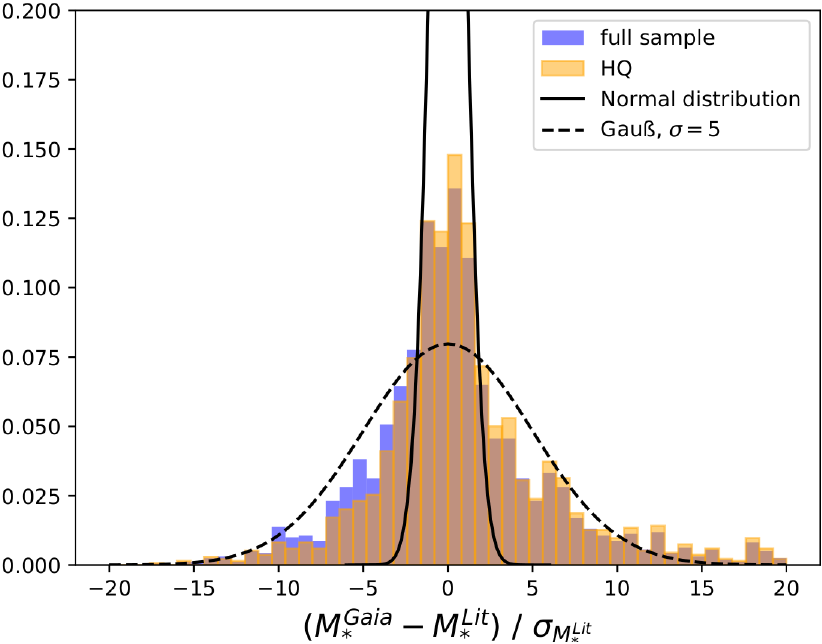} 
    \includegraphics[width=0.48\linewidth]{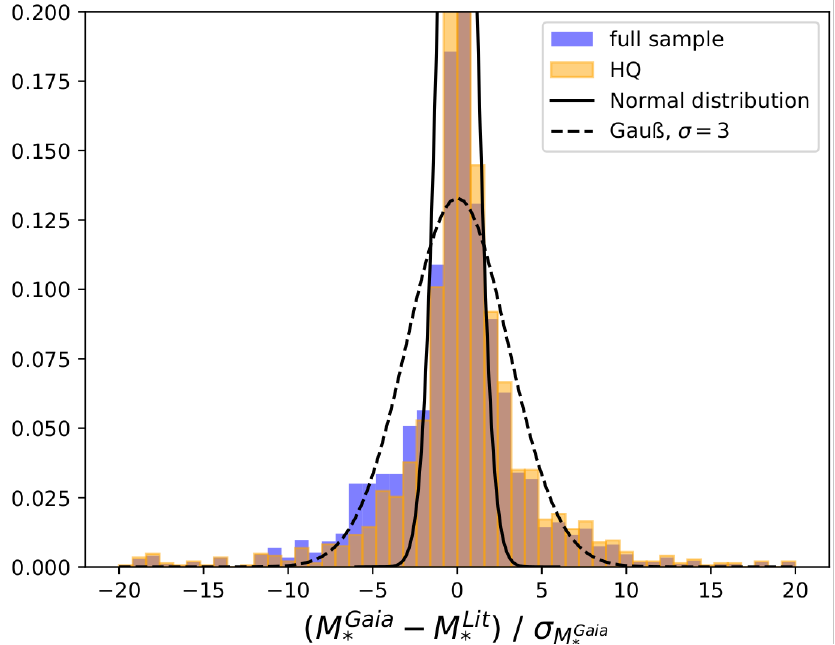} 
    \caption{Left panel: Same as Fig.~\ref{Fig:R_norm} for the stellar mass, but with a much broader range in abscissa. Right panel: same as left panel, but using the \Gaia\ uncertainty instead of the literature one.}
    \label{fig:ErrMass}
    \end{figure}

More importantly, when normalising the difference between the two catalogues by the uncertainty they provide, the distribution is very wide, with a standard deviation of the order of 13 with the \Gaia\ uncertainty, and seven with $\sigma_{M_\star}^{\Lit}$. In other words, each catalogue is totally inconsistent with the other, as illustrated in Fig.~\ref{fig:ErrMass}. Therefore, at least one of them underestimates tremendously its uncertainty, or is dramatically inaccurate. 

We notice in Fig.~\ref{Fig:M} that $M_\star^\Lit$ is strongly correlated to \T, while $M_\star^\Gaia$ much less.
We recall that $M_\star^\Gaia$ is derived independently from any stellar models, simply through $\log(g)$ and $R_\star$ (see Eq.\ref{eq:Mstar}). The former is almost independent of the estimation of \T, while the latter comes from $L_\star^\Gaia$ and  \T$^\Gaia$; in the end the link between $M_\star^\Gaia$ and \T$^\Gaia$\ is weak. In contrast, $M_\star^\Lit$ is probably estimated from isochrone fitting and stellar evolution models for most cases, which naturally produce a strong \T-$M_\star$ correlation. Finally, it is surprising to note that very few stars have $M_\star^\Lit \ga$1.5~$M_\odot$ (see also Fig.~\ref{Fig:ParamLit}).

We stress that the stellar mass is one of the most difficult parameters to obtain and we aim to show in Figs.~\ref{Fig:M} and \ref{fig:ErrMass}  that  these uncertainties ought to be taken with great care in future studies. In many cases,  only the internal error is provided, while the external error is actually much larger.

\subsection{Summary of the stellar parameter comparisons}
\label{Sec:CatalogConclu}

Comparing the stellar parameters derived from \Gaia\ and the ones found in the literature and downloaded from the online exoplanet tables, we find to first order a correct agreement (which is reassuring), except for the stellar mass with a standard deviation of $\sim 28\%$ for the $HQ$ sub-sample.
We also stress that the agreement is never perfect, and must conclude that different methods yield different results; the external error cannot be neglected. In summary, an analysis of the distributions shows that the \Gaia\ spectroscopic EHS catalogue is of high-quality, and provides smaller uncertainties than the literature. Hence, at least for the sub-sample with $HQ$ Flags, we recommend adopting this new catalogue, as explained in the  sections below.

\section{EHS population}
\label{Sec:EHSpopulation}

The Milky Way is known to be composed of a thin disc, defining  the Galactic plane and hosting ongoing star formation, along with a more diffuse thick disc, a central bulge, and a rather spherical and extended stellar halo.
Although these different Galactic populations can be intertwined and are rather difficult to disentangle\footnote{Several criteria could be adopted to identify these populations (chemistry, kinematics, dynamics,  and/or ages) but no consensus exists within Galactic experts. Following the discussions in \cite{PVP_Ale}, we adopt in Sect.~\ref{Sec:GalProp} the maximum height reached from the Galactic plane during their orbit to determine the population to which each EHS probably belongs.}, we could characterise the thin disc as composed by rather young metal-rich and low \AF\ stars with close to circular orbits. The thick disc is dominated by older, more metal-poor and \AF-richer stars having hotter dynamics. Finally, the Galactic halo mainly contains the oldest and most metal-poor \AF-rich Galactic stars, mostly with non-circular orbits. We refer, for instance, to \cite{PVP_Ale}, for a global view of the main characteristic of the Galactic components.
In this section, we analyse our EHS population using their physical and Galactic properties derived from \Gaia. The following section  is devoted to the exoplanets themselves.

\subsection{EHS atmospheric properties}
\begin{figure}[t]
    \centering
    \includegraphics[width=0.48\textwidth]{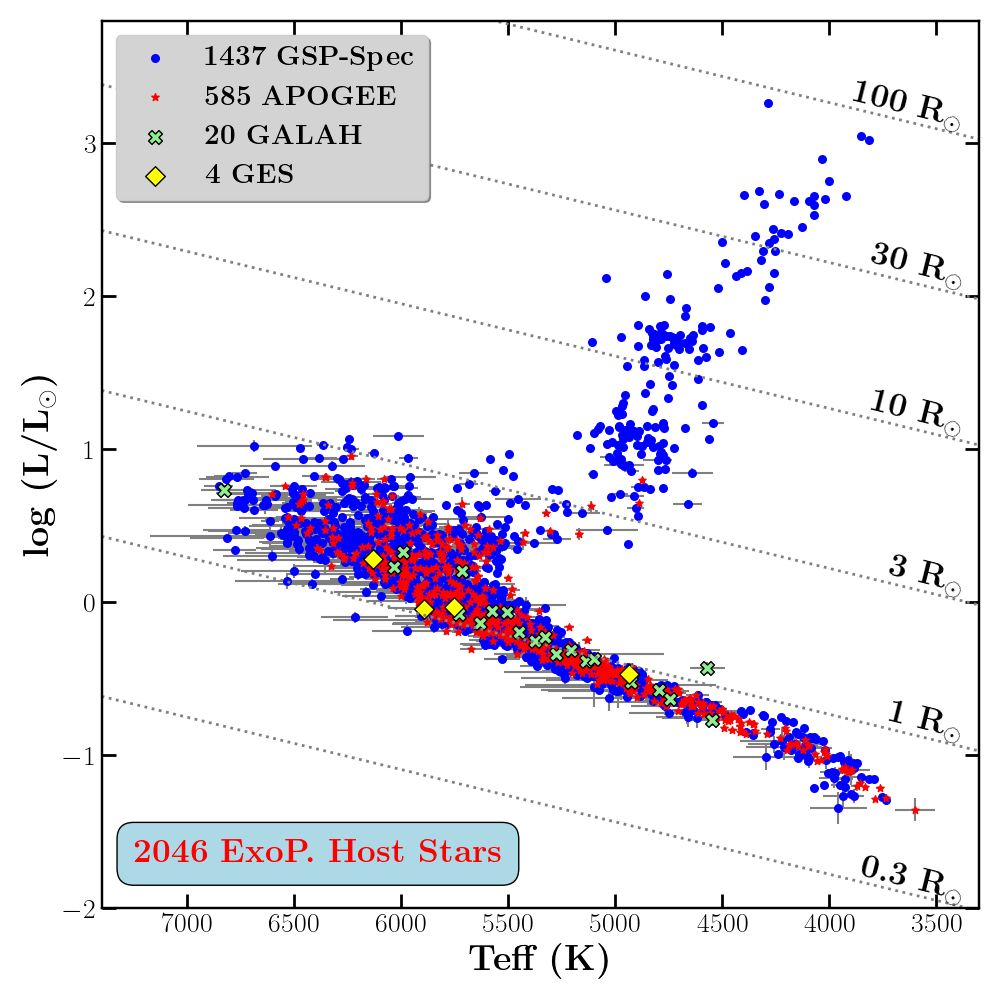} 
    \caption{Stellar luminosity versus effective temperature for the EHS sample. The different spectroscopic surveys that provided the atmospheric parameters are represented by specific symbols. Dotted lines represent the iso-radius relations.} 
    \label{Fig:L_Teff1}
\end{figure}

We illustrate some of the general properties of the \Gaia\ spectroscopic EHS catalogue by first showing in Fig.~\ref{Fig:L_Teff1} an effective temperature-luminosity diagram indicating from which spectroscopic survey the stellar parameters were adopted. We note that the \T\ uncertainties are clearly larger for dwarf stars than for giants since their parameterisation was performed from lower \SNR\ spectra. Hot dwarf EHSs have also larger temperature uncertainties than cooler dwarfs, illustrating that less and less lines are present in their spectra when \T\ increases. The large statistics and high homogeneity of  the EHS sample is also illustrated in Appendix~\ref{Appendix:L-T} and Fig.~\ref{Fig:L_Teff2}. 
In all these figures, we consider the whole catalogue described above without applying any flag and/or uncertainty filtering. 

\begin{figure}[h]
    \centering
    \includegraphics[width=0.48\textwidth]{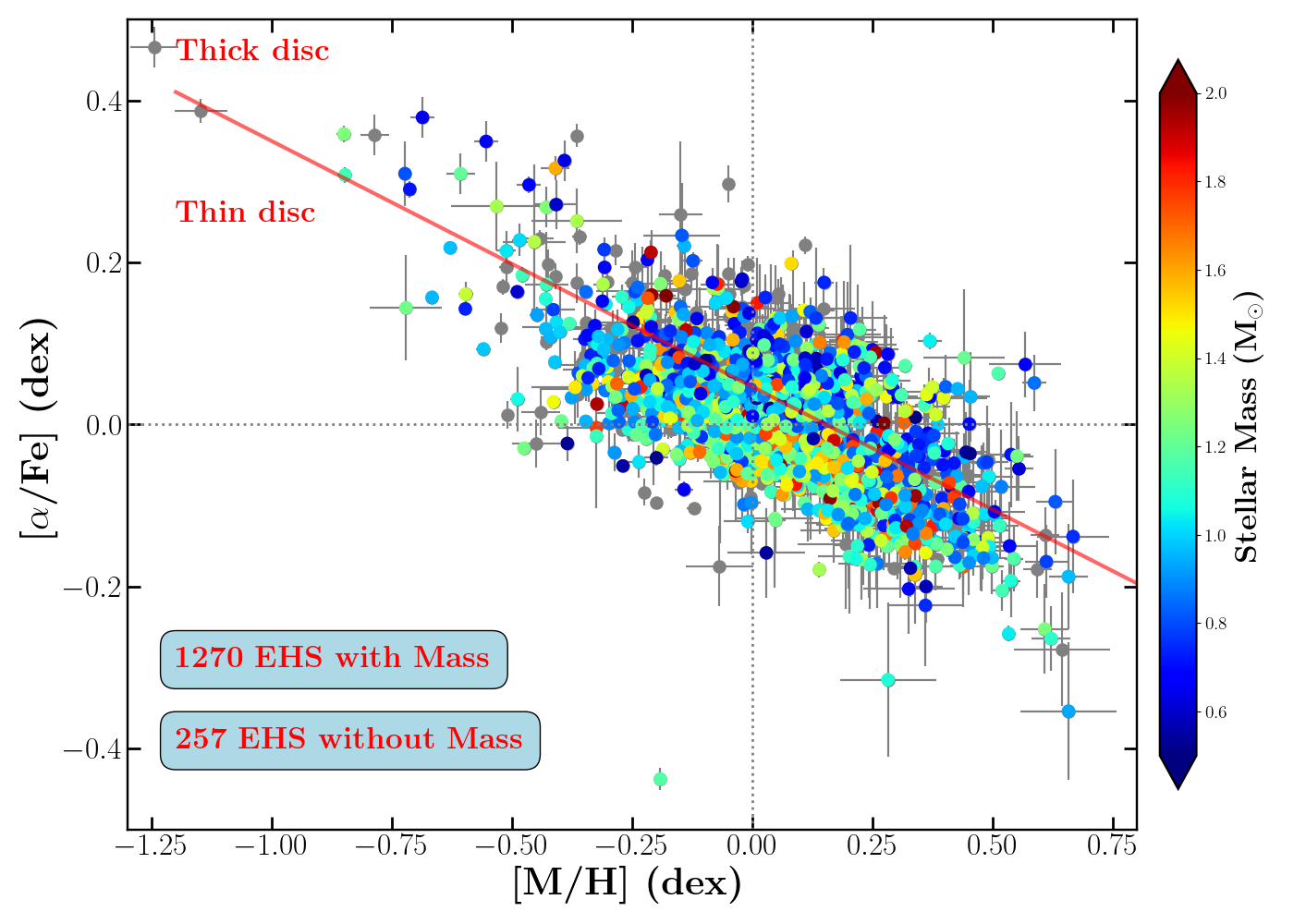} 
    \caption{Metallicity versus $\alpha$-elements over iron abundance ratio for the $HQ$ EHS sub-sample with high-quality chemical parametrisation (see text for details), colour-coded with their derived mass. The EHSs without published mass are plotted in grey. The Solar location is shown by the intersection of the dotted lines. The red line indicates a possible separation
between the thick (high-$\alpha$) and the thin (low-$\alpha$) discs based on chemical criteria from \cite[][see text for more details]{Pablo21}.} 
    \label{Fig:MetaAlpha}
\end{figure}

We show in Fig.~\ref{Fig:MetaAlpha} the chemical properties of the EHS $HQ$ sub-sample stars, filtering out stars with an uncertainty on \meta\ and \AF\ larger than 0.1~dex. 
Most of the EHSs are dwarf stars with masses and metallicities around Solar. Slightly more metal-rich EHSs are however found. The covered metallicity and mass ranges are quite large between [-1.2, +0.7]~dex and [0.5, 2]~$M_\odot$, respectively. This EHS sample is however not particularly metal-rich compared to Solar neighbourhood stars, as it was previously reported for giant-planet hosts \citep[see e.g.][]{Vardan19}.
Some red giant branch (RGB) or asymptotic giant branch (AGB) stars are also present, with the brightest EHSs being mostly rather metal-poor and slightly more massive than the Sun. 

The trend of \AF\ versus \meta\ is characteristic of the Galactic disc with a continuous decrease at super-Solar metallicities, as expected from Milky Way chemical evolution models \citep[see, for instance,][]{Ale_GES, Pablo21, PVP_Ale, Pedro23b}. 
As an illustration, we also show as a red line in Fig.~\ref{Fig:MetaAlpha} a possible separation between the thick (high-$\alpha$) and the thin (low-$\alpha$) discs,
as proposed by \cite{Pablo21} from the chemical analysis of AMBRE Project data \citep{AMBRE}. We note that this line was derived from the AMBRE [Mg/Fe] versus metallicity trend, which could not behave exactly as our \AF\ which is a proxy of \CaFe, another \alfa-element. Moreover, since Ca is not the easiest chemical species to chemically disentangle both discs \cite[see][]{PVP_Ale, Recio24}, we refer to Sect.~\ref{Sec:GalProp} for a discussion on the belonging of our EHS sample to the different Galactic populations based on their dynamical properties. 
Anyway, it can be seen that a large number of EHSs seem to have a low-$\alpha$ content typical of the thin disc but several others are more thick disc-like in terms of chemistry.
Finally, we note that a $\sim$1.2~M$_\odot$
star in Fig.~\ref{Fig:MetaAlpha} is very \AF-poor for its metallicity (\meta=-0.19 and \AF=-0.44).
We adopted the APOGEE atmospheric parameters for this star (Kepler-982, GDR3Id 2130393938771375104) that are fully consistent (in particular its metallicity) within error bars with those reported by SIMBAD at CDS. However, APOGEE reports \AF=0~dex and [Mg/Fe]=-0.1~dex for this star that are in agreement within each other and are more consistent with its metallicity, but are inconsistent with their published [Ca/Fe]. We therefore conclude that the APOGEE [Ca/Fe]  (which is our adopted \AF) should be revised for this star.

\bigskip

\begin{figure}[t!]
        \centering 
        \includegraphics[width=0.48\textwidth]{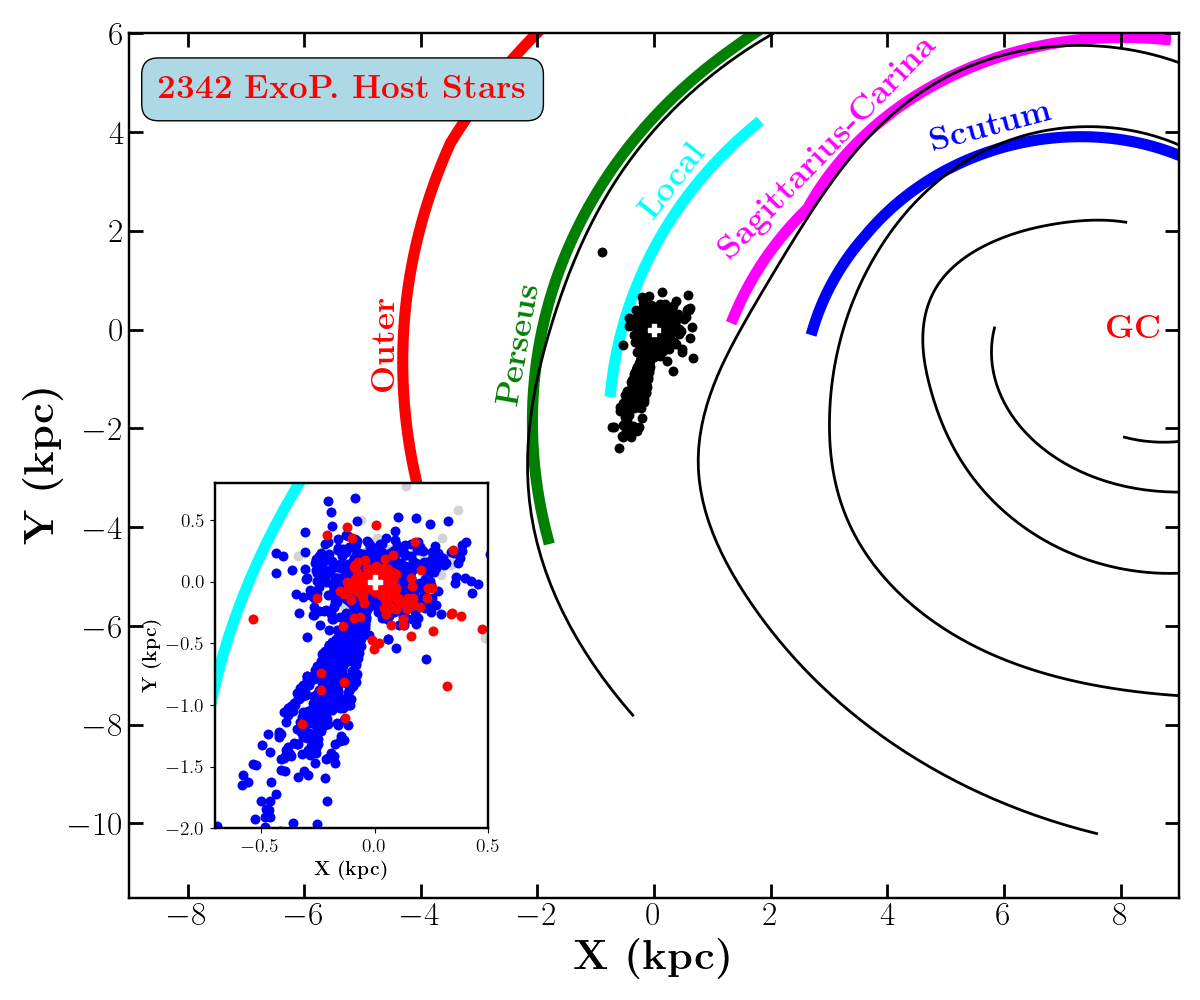}
        \caption{EHS sample in the Galactic disc. The spiral arms models of \cite{Georgelin76, Taylor93} and \cite{Reid19} are in black and in various colours, respectively. The Solar position is indicated by the white cross at (0,0) and the Galactic centre location is also shown. The inset is a zoom where most EHSs are found and its colour-code corresponds to giant (red) and dwarf (blue) stars, defined thanks to their radius (see text for details). EHSs without derived radius are shown in light grey in the inset.}
        \label{Fig:EHSinMW}
\end{figure}

\begin{figure*}[h!]
        \centering
        \includegraphics[scale = 0.22]{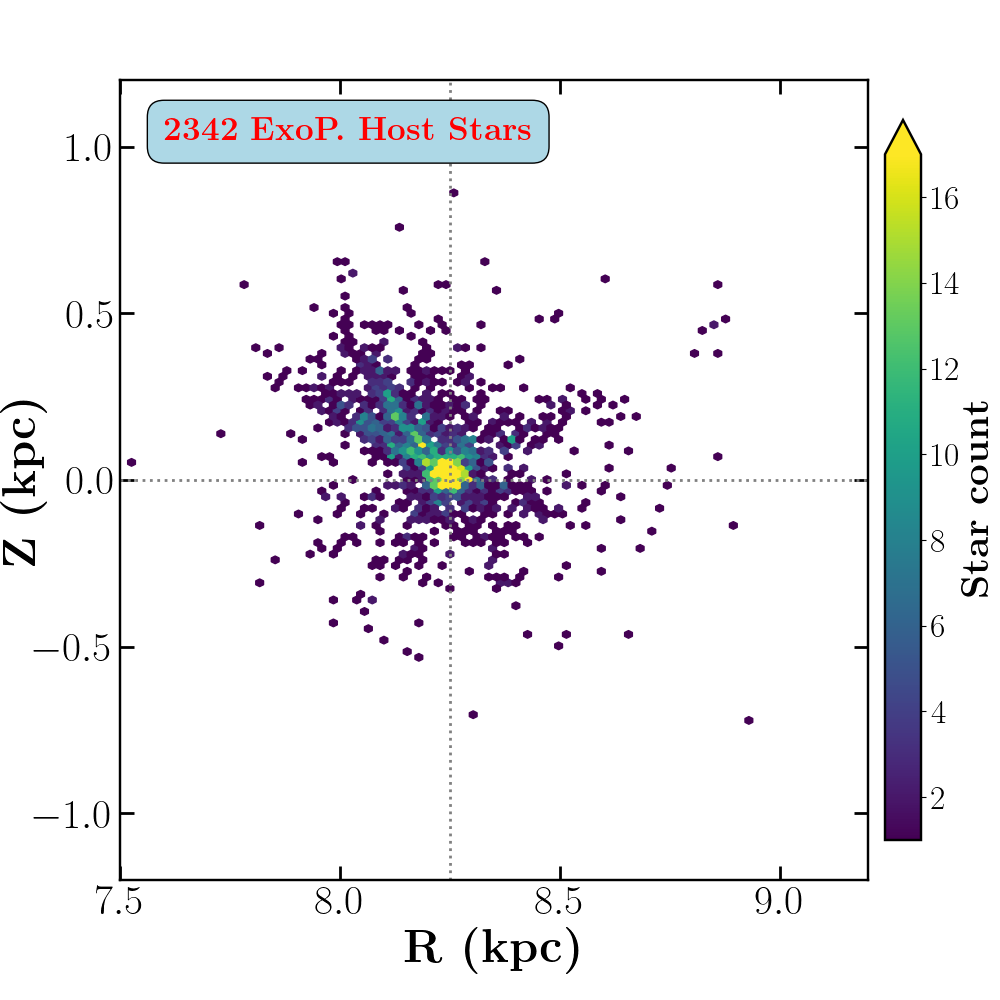}
        \includegraphics[scale = 0.22]{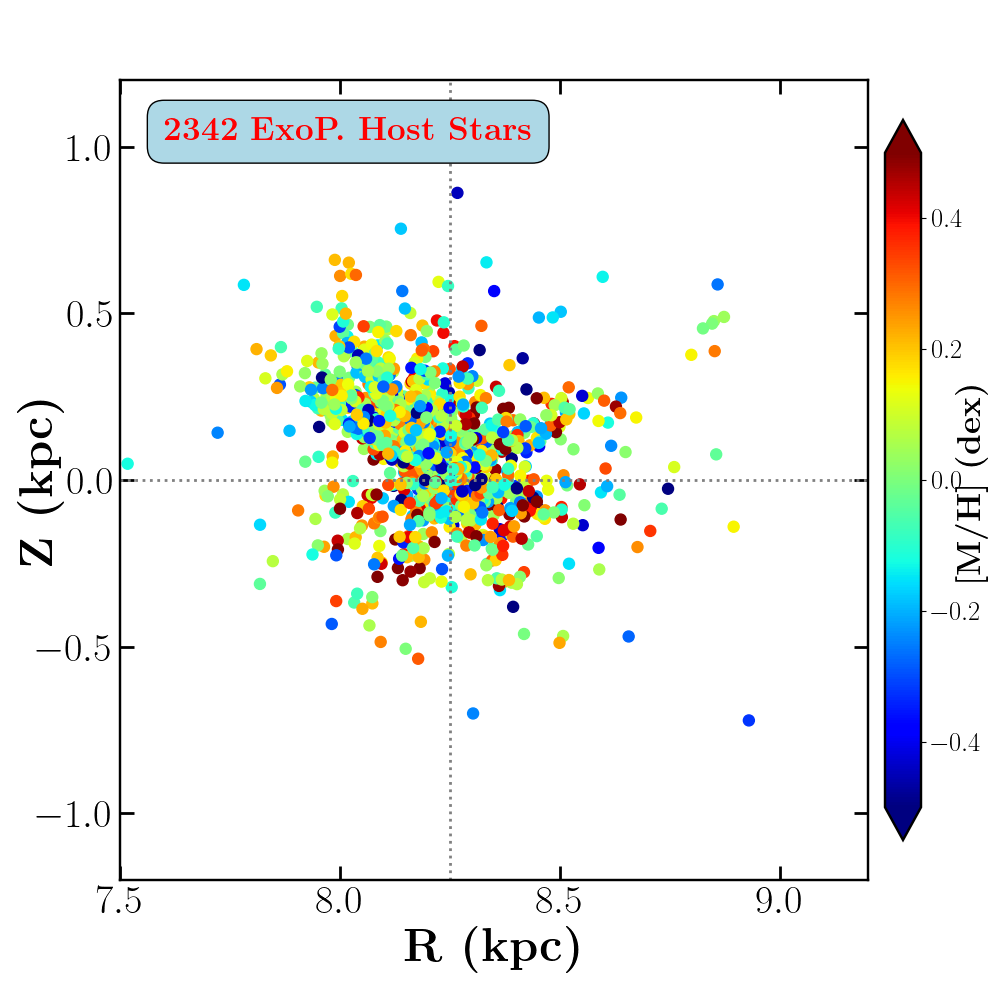}
        \includegraphics[scale = 0.22]{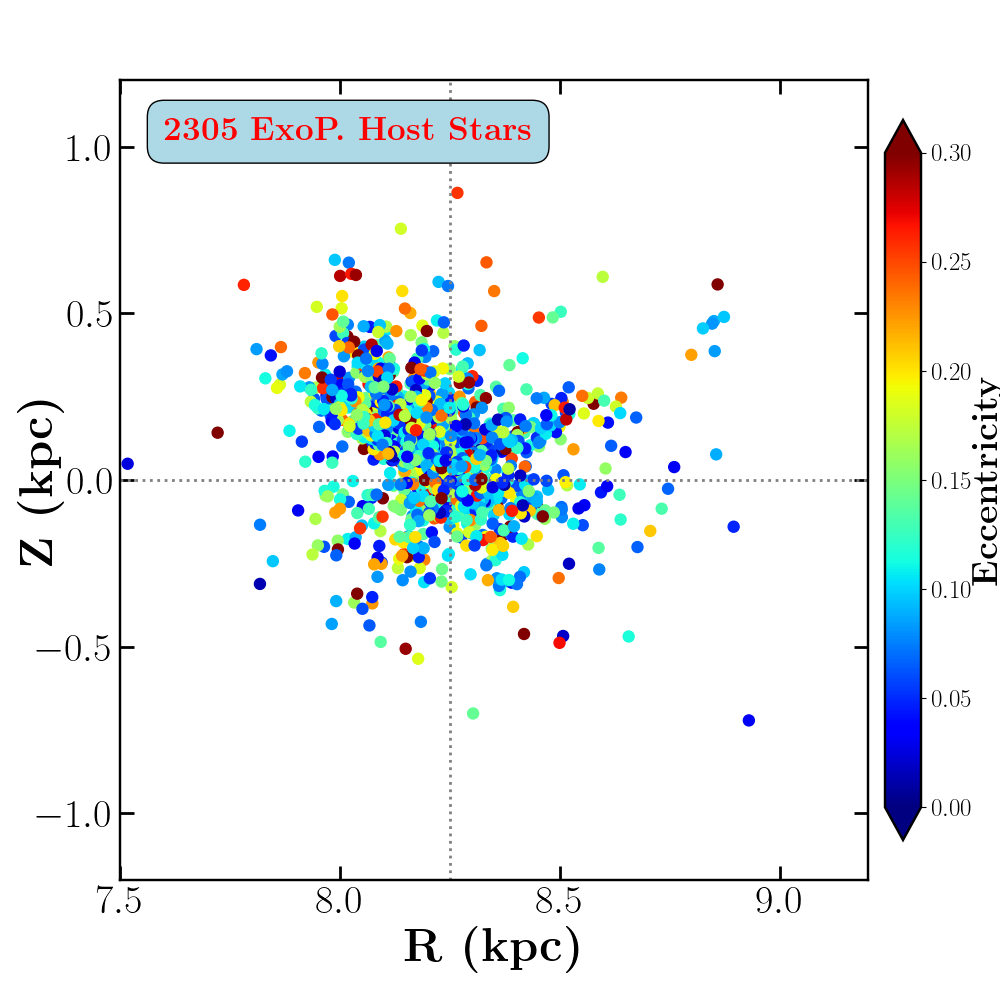}  
        \caption{Location of the EHSs in the Milky Way: Galactic distance above and below the Plane versus the Galactic centre distance. The Solar position is indicated by the intersection of the dotted lines. The colour-codes represent the stellar density, metallicity, and Galactic orbital eccentricity (from left to right, respectively).}
        \label{Fig:RZ}
\end{figure*}

\begin{figure}[h!]
        \centering
        \includegraphics[scale = 0.25]{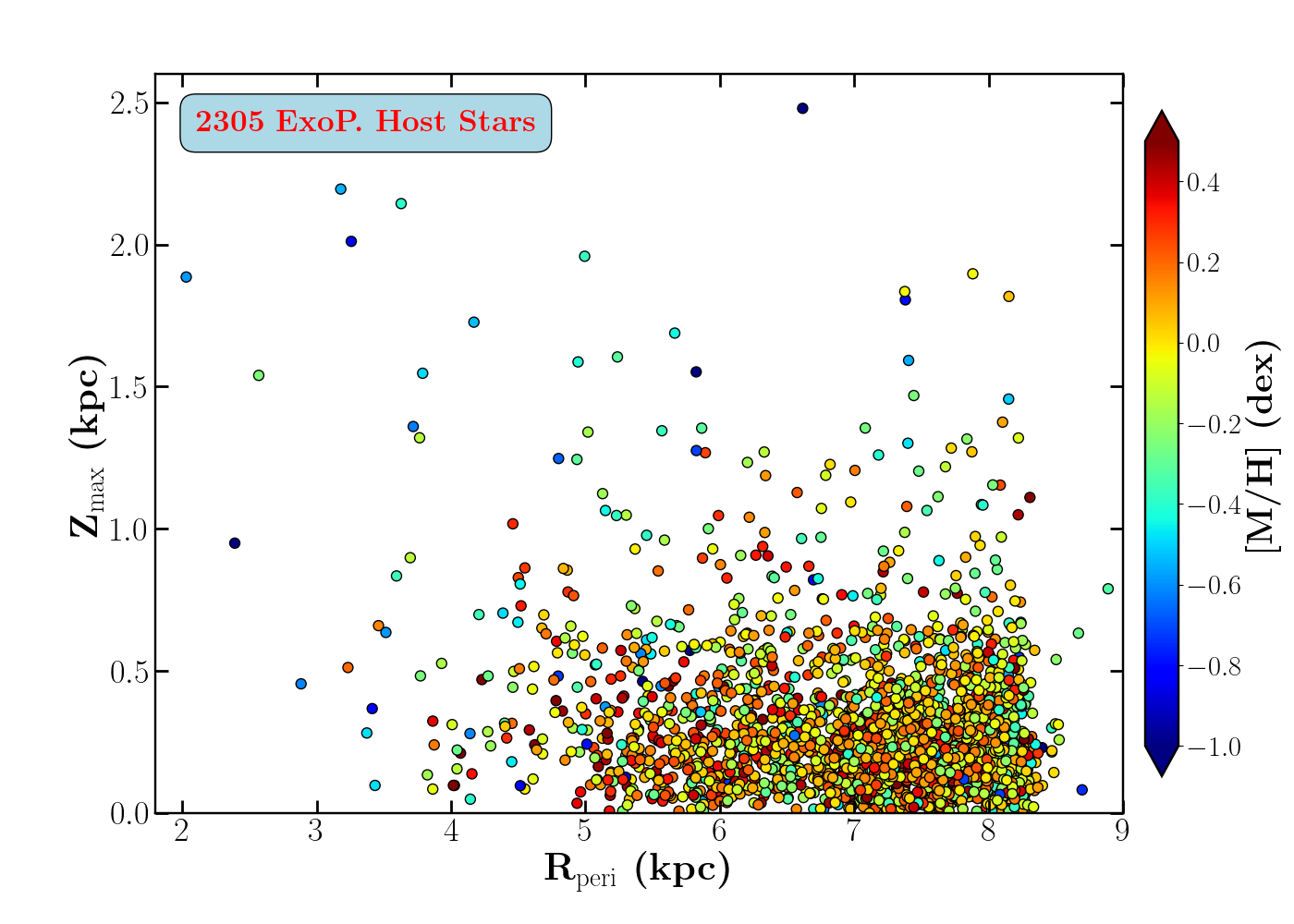}
        \caption{Orbital properties of the EHSs: largest distance above the Galactic plane during their orbit versus their pericentre Galactic distance, colour-coded with the metallicity. Some EHSs with eccentric orbits can reach high distances from the  plane, where the halo and thick disc populations are preferentially found. }
        \label{Fig:Rperi}
\end{figure}

\subsection{EHS Galactic properties}
\label{Sec:GalProp}
The location of the EHSs in the Milky Way is illustrated in Figs.~\ref{Fig:EHSinMW} and \ref{Fig:RZ}. We first show in Fig.~\ref{Fig:EHSinMW} where the EHSs are found in the Galactic plane, with respect to the main Galactic spiral arms whose precise location and nature is still a matter of debate \cite[see e.g.][]{Eloisa21, Marie25}. In this figure, the arm models in black are those of \citet[][]{Georgelin76, Taylor93}, derived from H~{\sc ii} regions; whereas the five arms of \citet{Reid19}, derived from the study of molecular masers associated with very young high-mass stars, are colour-coded. It can be seen from a Galactic point of view that all the exoplanet
presently detected can be found in the Solar vicinity, as expected, close to the Local arm. Therefore, most of the Galaxy exoplanetary content still needs to be explored. The large number of exoplanets detected by the $Kepler$ space mission is easily identified. The inset in Fig.~\ref{Fig:EHSinMW} shows the Solar neighbourhood where most exoplanets are detected and the colour-code refers to the EHS dwarf or giant nature. This dichotomy is based on the stellar radius, with giant stars in red defined as being higher than 3~$R_\odot$ (see also Figs.~\ref{Fig:L_Teff1} and \ref{Fig:L_Teff2}). Most of the closest exoplanets orbit around giant stars and they have typical masses close to the Jupiter one (see below the discussion on the planet masses).
Fig.~\ref{Fig:RZ} represents the EHS distance above or below the Galactic Plane versus their distance from the Galactic centre. The excess of EHSs towards the $Kepler$ field
is again clearly seen.
 Most of the other EHSs are located close to the Galactic plane with circular orbits and should thus belong to the thin disc. However, several stars are found at high distances from the Galactic Plane with rather eccentric orbits and low-metallicity. These stars are compatible to the Galactic thick disc or halo populations (see Appendix~\ref{Appendix:kinematics} for a discussion on their kinematics).

We  illustrate in Fig.~\ref{Fig:Rperi} the EHS Galactic orbital properties. 
Although most EHSs have circular orbits and stay in the Solar neighbourhood, it can again be seen that several other stars have eccentric orbits, leading them to reach high distances from the Galactic plane and/or the inner or outer parts of the Milky Way, where the halo and thick disc populations are preferentially found.
In this context, we have 392 EHSs (15\% of the whole sample) with an eccentricity larger than 0.2, $\sim$90 of them having $e$>0.3, i.e. a too large eccentricity to belong to the thin disc. These EHSs cover a very wide metallicity range
and have a pericentric distance indicating that they can cross or visit the most inner regions of the Milky Way disc during their life. We point out that six stars are not shown in this Fig.~\ref{Fig:Rperi} since their $Z_{\rm max}$ is larger than 2~kpc. Their metallicity is found within (-1.0, -0.25~dex). A few of them can have $Z_{\rm max}$ larger than 3~kpc
and can reach during their orbit the most external regions of the Milky Way (apocentre distance up to 11~kpc and $e$>0.45).  This suggests that they could be disc stars dynamically heated.

By adopting $Z_{\rm max}$\footnote{$Z_{\rm max}$ is available for 90\% of the whole EHS sample.
} to disentangle the different Galactic populations \cite[considering the criteria of ][]{Joss16}, we identify 56\% EHSs that probably belong to the thin disc ($Z_{\rm max}\le$300~pc). About 31\% are found in the thick disc ($Z_{\rm max}$ within 0.3 and 1.0~kpc) and almost 3\% are probably member of the Galactic halo ($Z_{\rm max}$>1.0~kpc). Adopting $Z_{\rm max}$ as a separation criteria could be discussed but we recall that following for instance \cite{PVP_Ale}, a good correlation exist between this quantity and the disc chemical properties. We can also refer to \cite{Pedro23} who adopted a similar criteria to define their thin disc sample. Nevertheless, we again point out that other adopted criteria to define the Galactic population membership (as chemistry, kinematics, ages, ...) could lead to rather different statistics, and thus, conclusions \citep[see e.g.][and footnote\#12 of the present article]{Swastik23}.

The two extreme EHS halo member with $Z_{\rm max}$>3~kpc are GDR3Id 790236569006916096 (HD-233832) and GDR3Id 213097678731141235 (Kepler-1578). They are not extremely metal-poor for belonging to the halo (\meta=-0.69 and -0.27~dex, respectively) but are $\alpha$-enriched as expected
(\AF=0.38 and 0.27~dex). They orbit around the Galactic Centre with an eccentricity of 0.26 and 0.48, respectively. Finally, we find that they have rather low masses (0.65 and 0.8~M$_\odot$, respectively) and, therefore, they could be rather old, but they do host exoplanets.
We also emphasise two interesting metal-poor stars 
which quality flags are less than two (i.e. a good parameterisation) and with peculiar kinematics properties: HD11755 (GDR3Id 558504529130235136, a cool giant with \meta=-1.0~dex) and 
HD155918 (GDR3Id 5801950515627094400, a Solar-type star with \meta=-0.8~dex and \AF=+0.3~dex), both being located in the Solar vicinity. Their kinematics and orbital properties ($Z_{\rm max}$=2.5 and 2.0~kpc,  $e$=0.25, and 0.44, respectively) reveal that they probably belong to the Galactic halo, and, again should therefore be rather old.
Rather similar conclusions can be derived for the cool 
main sequence star HD155358 (GDR3Id 1334643739861016832 that hosts two planets) with $Z_{\rm max}$=0.6~kpc and $e$=0.19, revealing that it probably belongs to the thick disc. However, we note that our reported metallicity (\meta=-1.15~dex) for this star could be lower by $\sim$0.5~dex than some values reported in CDS/Simbad. 
Finally, HD80913 (GDR3Id 5195752440555653504) characterised by \meta=-1.2~dex and \AF =0.6~dex (but with uncertainties around 0.2~dex) is orbiting in the Galactic thin disc, but with a large eccentricity close to 0.26.\\

Finally, we have shown that some EHSs belong to the oldest Galactic components: the thick disc and the halo.
We recall that these two populations are known to be composed by old stars: typical ages for thick disc stars are older than $\sim8$~Gyr \cite[see, for instance,][]{Michael17, Pablo21, Rix22} whereas halo stars could be even older. 
 The detection of exoplanets already formed at the earliest epochs in the Galactic history is therefore  very interesting for constraining the planet formation in diverse environments,
 that can be extreme in chemistry, kinematics, and/or dynamics properties.
Several of these EHSs should deserve further studies to explore the nature of their exoplanet companions and the formation of such systems in the oldest Galactic populations.

\section{Exoplanet properties}
\label{Sec:Planets}

\subsection{The \Gaia\ spectroscopic catalogue of exoplanets}
\label{sec:Gaia_cat_pl}

From the above calculated new EHS properties (see Table~\ref{Tab:EHS}), it is then possible to derive new exoplanetary parameters. To build this exoplanetary catalogue, we proceeded as follows, after having first defined a specific flag ($Flag_{\rm source}$) in order to easily identify the exoplanets table from which the data were recovered. We first considered the exoplanets in the NEA table, by default, which correspond to $Flag_{\rm source}$ = 1 (277 exoplanets, after applying the other filters defined below). 
For each of them, we then checked if it is also in the EES table. If so, we verified the compatibility between the two data sets, by computing in each table the ratio of planetary to stellar radii and/or masses taking into account the uncertainties. If these ratios are compatible 
(that is: the $1\sigma$ error bars for the value of the ratios in both tables partly overlap),
we then kept the NEA values and set $Flag_{\rm source}$ = 0 (4837 exoplanets). If not, we still kept the NEA values but set $Flag_{\rm source}$ = 3 (40 exoplanets). Finally, we added all the other exoplanets, which are only found in the EES table, and gave them a $Flag_{\rm source}$ = 2 (1330 exoplanets). 
We note that the online databases provide only one value for the planet radius or mass, even if several observations of the given system can be found in the literature.

In the end, we also checked the consistency of each entry in Tab.~\ref{Tab:ExoP} by recomputing the planetary masses (resp. radii) using the stellar mass (resp. radius) and the orbital and radial velocity (resp. transit) parameters\,; if this mass (resp. radius) differs from the one given directly in the table by more than $10\%$, we set $Flag_{\rm source}$ = 9 (278 cases, with 56 being inconsistent for the radius and 223 for the mass --\,hence, one for both).

This list of exoplanets was then merged with the \Gaia\  EHS catalogue to retrieve the \Gaia\ spectroscopic parameters of the corresponding stellar hosts and then to derive the new planetary radii (\RpGaia) and masses (\MpGaia~$\sin i$). In doing so, we combined the columns giving the planet mass and the planet minimum mass, $M_p\cdot\sin(i)$\,: if both are provided, we kept $M_p$ by default, but we checked that $M_p$ is truly equal to $M_p\cdot\sin(i)/\sin(i)$ when $i$ is provided, and if not (more than $10\%$ difference), we set $Flag_{\rm source}$ to 9 (with nine cases with already $Flag_{\rm source}=9$ and three new cases).
This leads to 2573 stars (1826 being $HQ$) for 3556 exoplanets for the full sample. Finally, we hereafter label $HQ$ a planet whose EHS is $HQ$ and has a $Flag_{\rm source}<9$; there are 2346 such $HQ$ exoplanets in our catalogue.

\paragraph{Planetary radius rescaling}
Planetary radii are measured indirectly with the transit depth, which is the square of the planet to star radius ratio. Hence, $R_p\propto R_\star$. Thus, whenever the planet detection method or the radius determination method is said to be 'primary transit' in the online table, we simply rescaled the planetary radius by the stellar radius:
\begin{equation*}
R_{\rm p}^{\rm Gaia} = R_{\rm p}^{\rm Lit} \times \left(\frac{R_\star^{\rm Gaia}}{R_\star^{\rm Lit}}\right)\ .
\end{equation*}
This means that, unfortunately, if we cannot estimate \RstarGaia, then \RpGaia\ cannot be obtained either. There are 19 planets in the table that have a reported radius, which have been detected through direct imaging. In such cases, the planetary radius was estimated independently from the stellar radius and we did not change it: \RpGaia = \RpLitt.

There are also six planets in the table that have a radius, but have been detected by radial velocity (or some unknown method) and do not transit their host star. In this case, the planetary radius is probably derived from theoretical mass-radius relations\,; we chose to remove these radii from the table because we wanted to observationally constrain the mass-radius diagram (see the next section).

\paragraph{Planetary mass and semi-major axis rescaling}
Similarly, when a planet mass is estimated using the radial velocity method, astrometry, or transit timing variation, we re-scaled the planetary masses and semi-major axes of the literature by (\MstarGaia/\MstarLitt)$^{2/3}$, and (\MstarGaia/\MstarLitt)$^{1/3}$, respectively.

As for the radius, we found planets with \MpLitt\ that have been detected by direct imaging (83 cases). In this case, we did not rescale their mass since this method yields a planetary mass independently of the host star. Nonetheless, we have provided the stellar parameters from the \Gaia\ estimate, as explained in the previous section. 
In addition, 70 planets have a mass although they have been detected by transit and no indication of their mass estimation method is given. In a planetary mass-radius plot, such as that shown in Fig.~\ref{Fig:Mp-Rphist}, we can see that 65 of these 70 planets that also have a radius are randomly scattered and clearly do not follow the general trend. Hence, we considered these masses to be very suspicious and we removed these 70 masses from our table.
It is worth noting that this cleans the region $M_p>0.3\,M_{\rm Jup}$, $R_p<0.7\,R_{\rm Jup}$, where we can find 21 planets in the tables downloaded from the literature, whose large density is hard to explain. Only five of them still have a mass in our catalogue, two of them being henceforth outside of this zone thanks to the re-scaling (see Fig.~\ref{Fig:Mp-Rphist} below).

\paragraph{Calculation of the errors}
We cannot estimate the uncertainty on \RpGaia\ and \MpGaia\ by a direct propagation of errors because the error on the transit depth or radial velocity amplitude is not always provided. Instead, knowing that $\epsilon($\Rp$)^2 = \epsilon($\Rstar$)^2 + \epsilon(\sqrt{TD})^2$ 
(where $TD$ is the transit depth and $\epsilon(Q)$ denotes the relative uncertainty associated with $Q$\,: $\sigma_Q/Q$), we first estimate $\epsilon( \sqrt{TD})^2 = \epsilon($\Rp$)^2 - \epsilon($\Rstar$)^2$ in the table and check if it is positive. If not (which is an indication of a problem in the data given it is not possible for the planetary radius to be better known than the stellar radius), we set $\epsilon(\sqrt{TD})=0$  and thus use $\epsilon($\Rp$) = \epsilon$(\RstarGaia), otherwise we keep the full value $\epsilon($\Rp$)^2 = \epsilon($\Rstar$)^2 + \epsilon(\sqrt{TD})^2$. We proceed similarly with the planetary mass and associated measurements used to obtain it, which are all included in $\epsilon($\Mp$)^2-\big(\frac23\epsilon($\Mstar$)\big)^2$. 

We stress that we found no less than 494 cases for which the relative error on the planetary radius is smaller than that on the stellar radius in the table, and 1157 for which the error on the planetary mass is smaller than two thirds of that on the stellar mass. This does not make sense, unless the error on the radii and masses have been computed taking only into account the uncertainty in the transit depth and radial velocity measurements, neglecting the uncertainty in the stellar parameters. We are afraid that these numbers show that this is often done. We point out that as a consequence, although our stellar parameters are most often better constrained than in the literature (see previous section) there are many cases where the uncertainty on the planetary parameter is larger in our catalogue than in the literature. But this is only due to the fact that we do take the uncertainty on the stellar parameter properly into account.

\begin{table}[t]
        \caption{\label{Tab:ExoP} 
        \Gaia\ spectroscopic catalogue of exoplanet properties.}
        \centering
        \begin{tabular}{ll}
        \hline
        Label & Description \\
        \hline
        GDR3Id & $Gaia$ DR3 Identification\\
        Planet name &  Planet name in the EHS databases\\
        \MpGaia $\sin{i}$ & Planetary min. mass computed from \MstarGaia\ (M$_{\rm Jup}$)\\
         \MpGaia $\sin i_{\rm err}$ & Associated uncertainty to \MpGaia $\sin i$\  (M$_{\rm Jup}$)\\
        \RpGaia & Planetary radius computed from \RstarGaia\ (R$_{\rm Jup}$)\\
        \RpGaia$_{\rm err}$  & Associated uncertainty to \RpGaia $\sin{i}$ (R$_{\rm Jup})$ \\   
        $a$ & Rescaled semi-major axis (AU)\\
        $a_{\rm err}$ & Uncertainty on $a$ (AU)\\
        $K$ & RV semi-amplitude from Exop. catalogue (m$\cdot s^{-1}$)\\
        $K_{\rm err}$ & Uncertainty on $K$ (m$\cdot s^{-1}$)\\
        $e$ & Eccentricity from Exoplanet catalogue \\
        $i$ & Inclination of the system (Exop. catalogue, deg) \\
        $TD$ & Transit depth from Exoplanet catalogue ($\%$) \\
        $TD_{\rm err}$ & Uncertainty on $TD$ ($\%$) \\
        $P$ & Orbital period from Exop. catalogue (days)\\
        Flag$_{\rm source}$ & Source of planetary parameters and \\
                            & compatibility between the tables. \\ &0: NEA compatible with EES (2834 planets); \\ &1: NEA only (121 planets);  \\ 
                            &2: EES only (367 planets); \\
                            & 3: NEA not compatible with EES (28 planets); \\
                            & 9: Inconsistency within the tables (206 planets).\\
         \hline
        \end{tabular} 
        \tablefoot{The full version of this table is available in electronic form at the CDS.}
\end{table}

\subsection{Exoplanet mass-radius distribution}
\label{Sec:MassRadDist}

\begin{figure}[t]
        \centering
        \includegraphics[width=\linewidth]{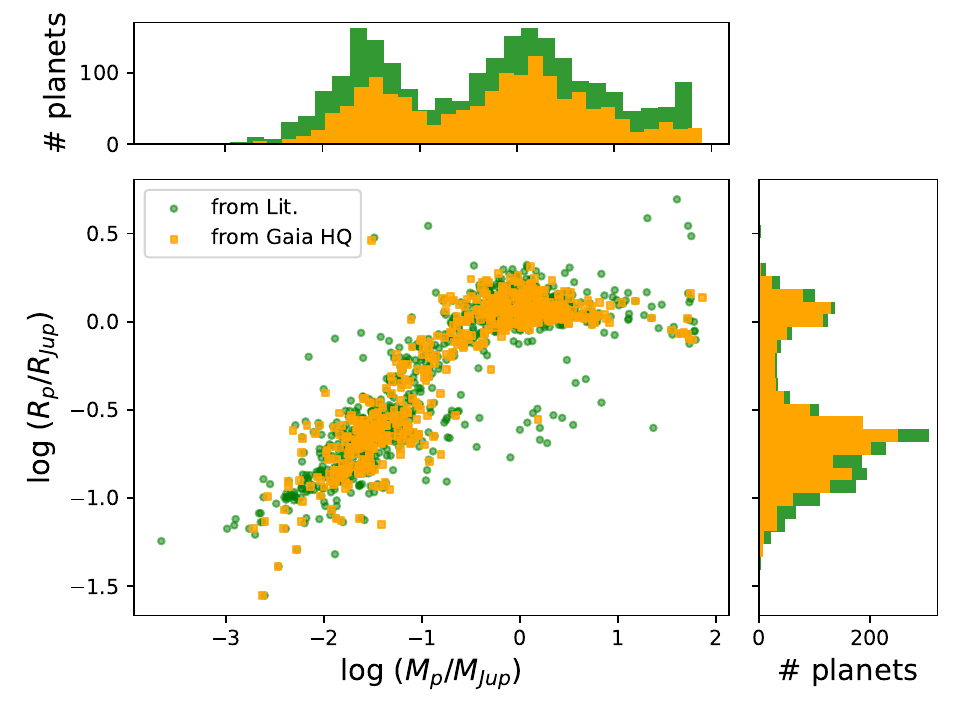}
        \caption{\Mp\ vs \Rp\ in log scale, with the corresponding distributions for exoplanets found in the literature (green) and those of the $HQ$ sample (orange) for which the exoplanets properties were recomputed from \MstarGaia\ and \RstarGaia.}
        \label{Fig:Mp-Rphist}
\end{figure}

The \Mp --\Rp\ distribution from the literature and from our new catalogue ($HQ$ sample) is shown on Fig.~\ref{Fig:Mp-Rphist}.  Since the \Gaia\ exoplanet radii and masses are computed from  \MstarGaia\ and \RstarGaia, the mean, median and standard deviation of the planetary parameters ratios between the literature and our catalogue are similar to the stellar ones. 
We retrieve the known distributions of exoplanets (in the scatter plot and in the histograms), meaning that the rescaling of the exoplanetary radii and masses do not affect the general population of exoplanets.
This is not surprising since the planetary radii (resp masses) span more than one (respectively, four) orders of magnitude, while the EHS radii (resp masses) generally vary by only $\sim 10\%$ ($\sim30\%$).

\begin{figure}
        \centering
        \includegraphics[width=\linewidth]{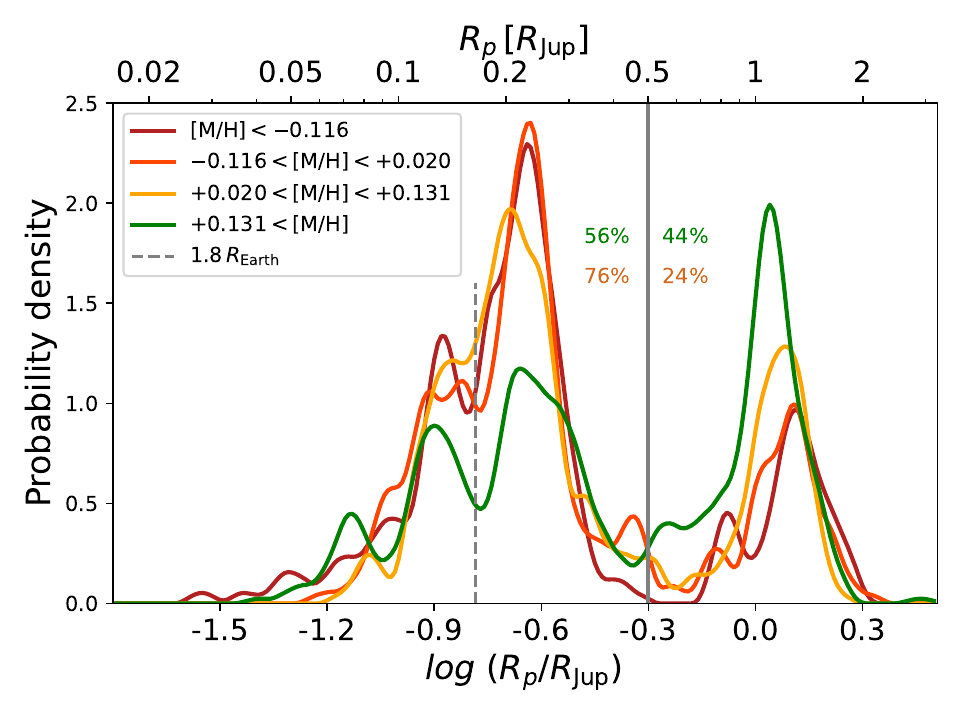}
        \caption{Distribution of exoplanets radius
        for four different bins in \meta\ of their host stars. The probability density is obtained using a Kernel Density Estimate with a $\cos^2$ kernel of width $0.05$~dex.}
        \label{Fig:Hist_Rp-meta}
\end{figure}

We now separate our $HQ$ sample of exoplanets adopting the same bins in metallicity of the stellar host than \cite{Petigura2018}, i.e. four bins with \meta<-0.116~dex, -0.116<\meta<+0.020~dex, +0.02<\meta<+0.131~dex, and \meta>+0.131~dex 
that they chose to obtain the same number of stars in each of their bin. We restrict our sample to planets that have a radius, that is 1290 exoplanets. This yields 242, 275, 246 and 535 exoplanets in each bin respectively, whose radius distribution is shown in Fig.~\ref{Fig:Hist_Rp-meta}. We notice a clear trend where $44.4\%$ of the planets found around stars with \meta>0.131~dex are gas giants (\Rp$>0.5$\Rjup) but only $28$, $23$, and $21\%$ in the other bins in decreasing order of metallicity, for a total of $24\%$ for EHS with \meta<0.131~dex.
It is actually well established in the literature that the occurrence rate of giant planets increases strongly with stellar metallicity \citep[e.g.][]{Santos01, Johnson10,Buchhave2012,Petigura2018}, although a larger scatter is seen above $\sim4$~\Mjup\  \citep{Buchhave14, Santos17, Swastik21}. Moreover, a wider range of metallicities is found for a large variety of exoplanets.

\subsection{The radius valley}
\label{sec:Fulton}
\begin{figure}[t]
        \centering
\includegraphics[width=\linewidth]{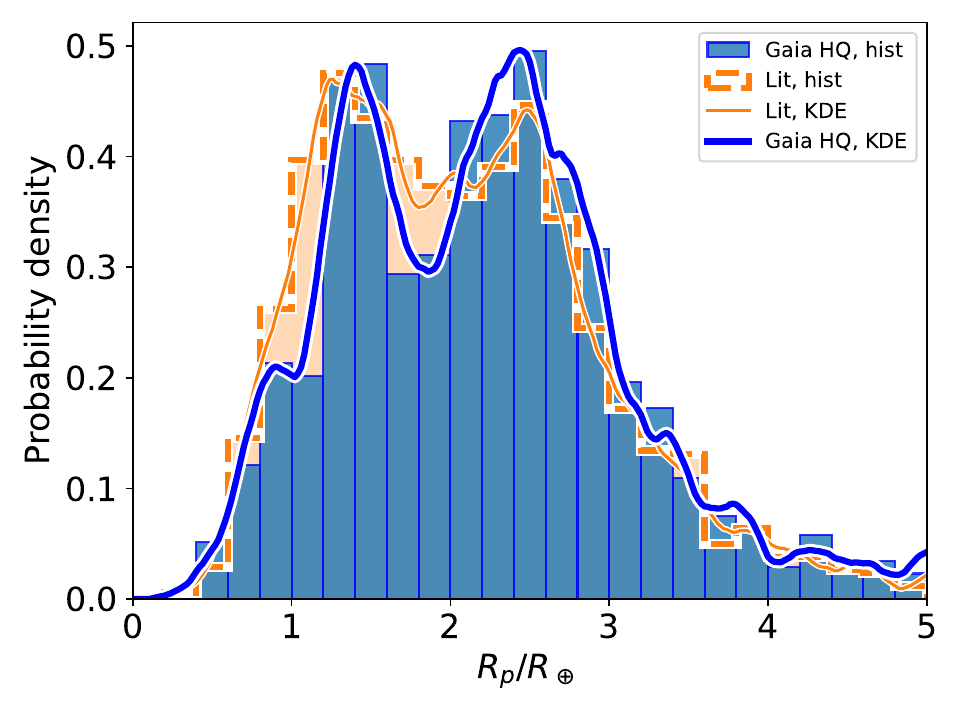}
\caption{Histogram of the non-giant exoplanets radii, with bins of $0.2\,R_\oplus$ and Kernel Density Estimation using an Epanechnikov (parabolic) kernel of width $0.2\ R_\oplus$. Orange: Literature. Blue: Our \Gaia\ $HQ$ sample. The so-called radius valley around $\sim1.8\,R_\oplus$ is more prominent in our $HQ$ sample.}
        \label{Fig:histRp_zoom}
\end{figure}

The division between giant planets and non-giants is very clear on Fig.~\ref{Fig:Hist_Rp-meta}, around \Rp$~\approx 0.5$\Rjup$\,\approx 5\,R_\oplus$.
Focusing now on non-giants (\Rp$<5\,R_\oplus$), Figure~\ref{Fig:histRp_zoom} shows  
in blue the histogram of the planetary radii in our $HQ$ sample, together with its Kernel Density Estimate using an Epanechnikov kernel of half-width $0.2~R_\oplus$ (bold blue line). 
We recover
a bimodal distribution with a separation at $\sim$1.8~$R_\oplus$ (vertical dashed line in Fig.~\ref{Fig:Hist_Rp-meta}), as first noticed by \citet{Fulton17}. We note that this gap appears significantly more marked in our $HQ$ sample than using the literature values (in orange). More precisely, $V_A$\footnote{\label{FootVa}Defined by \cite{Fulton17} as the ratio of the number of planets with $1.64<R_p/R_\oplus<1.97$ to the square root of the product of the number of planets with $R_p/R_\oplus$ between $1.2$ and $1.44$ and between $2.16$ and $2.62$.} is $0.83$ for the literature against $0.67$ in our $HQ$ sample, where $1$ corresponds to a flat distribution in logscale and a smaller value of $V_A$ corresponds to a deeper valley.
This is consistent with the gap being a real feature which is blurred by the uncertainty in the planetary radii caused by the uncertainty in the stellar radii. In this case, since the \Gaia\ stellar radii are more accurate as shown in Sect.~\ref{Sec:Catalog}, the gap is less blurred, as observed. Figure~\ref{Fig:histRp_zoom} therefore confirms the existence of a radius valley, which the accuracy of the parameters provided in our catalogue allows us to get a better glimpse of the data.

\begin{figure}
        \centering
        \includegraphics[width=\linewidth]{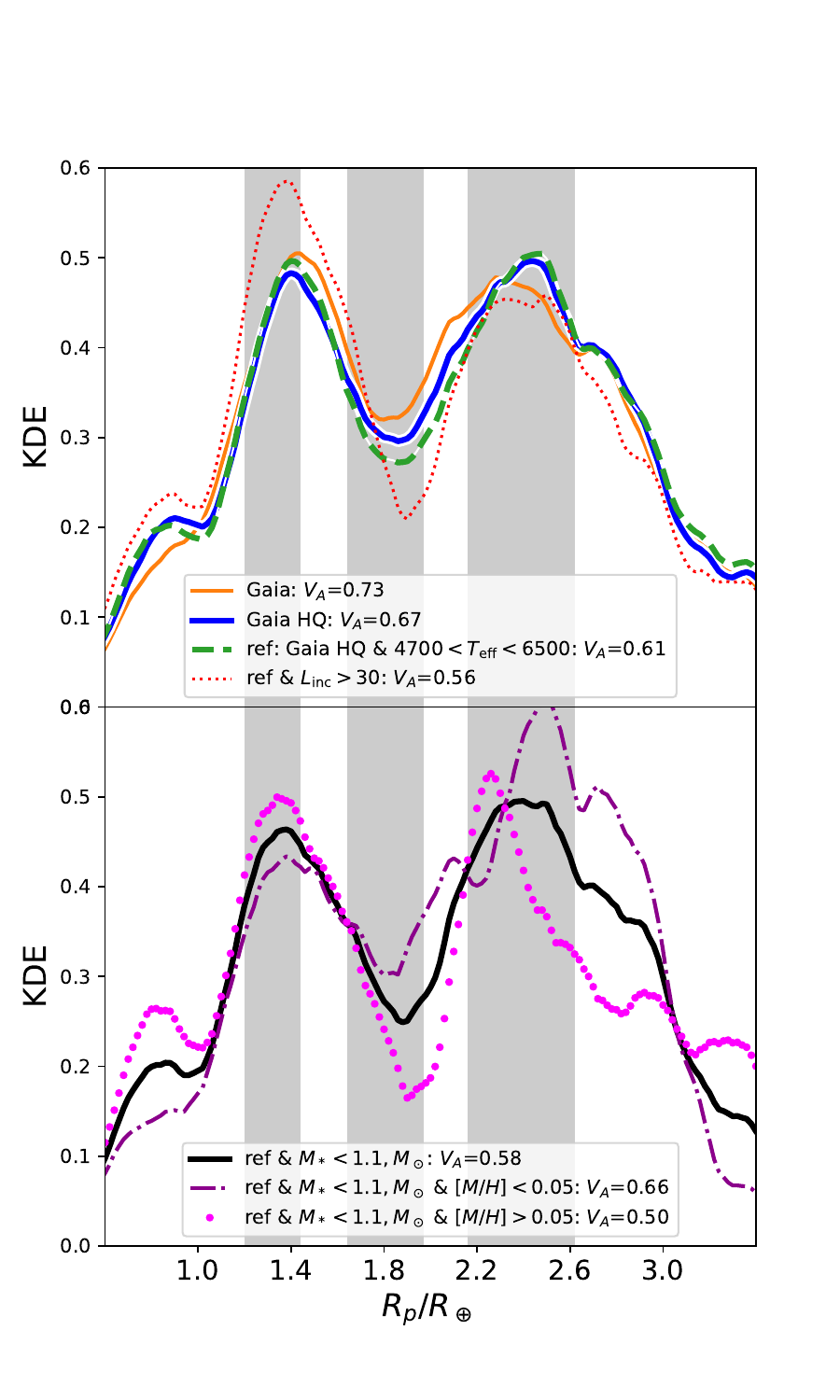}
        \caption{KDE of the radius distribution of exoplanets of $R_p<5R_\oplus$ for different sub-samples, zoomed around the radius valley. Top panel: Our whole catalogue (orange), the $HQ$ sample (blue, same line as in Fig.~\ref{Fig:histRp_zoom}), $HQ$ sample with host star effective temperature between 4700 and 6500~K (green dashed, reference sample), and reference sample restricted to planets suffering more than 30 times the stellar incoming luminosity received by the Earth. Bottom panel: Reference sample for $M_*<1.1\,M_\odot$, split in two bins of metallicity of the host star. Grey regions: Radius ranges used to compute $V_A$ (see footnote~\ref{FootVa}).}
        
        \label{Fig:FG1}
\end{figure}

From this figure and following \citet{Fulton17}, we explored the KDE of various exoplanets samples in the region of interest (Fig. \ref{Fig:FG1}), where the intervals mentioned in footnote~\ref{FootVa} appear in grey (they are ranges of equal width in logscale: $0.08$~dex).
This allows us to estimate how the shape, width and depth of the gap depends on the various parameters. We note that the shape of the curves also depends on the chosen kernel, and that they would look slightly different if we make the KDE on log(\Rp) instead of \Rp;
we also see from the curves that the value of $V_A$ may depend on the reference ranges chosen. We checked to make sure this did not affect our results.

We find (see top panel) the fact that the radius valley is more marked if we restrict the sample to exoplanets whose host star effective temperature is between 4700 and 6500~K -- and even more for the ones receiving an incoming luminosity 30 times greater than that of the Earth from their
respective stars. We also see that $1.8\,R_\oplus$ is a good estimate for the gap in all samples, although a slightly larger value may be even better, especially for the last sample.
In the bottom panel, we split our reference sample ($HQ$ and $4700<T_{{\rm eff},*}<6500$\,K) into two bins of host star metallicity.
However, since massive stars are short-lived and metal-poor stars are formed early in the Galactic history, there exists no massive metal-poor stars: for instance, in our $HQ$ sample, we have no star with \meta<-0.2~dex and \MstarGaia>1.1~$M_\odot$. This introduces a bias, which we removed by applying an additional filter and discard all EHS more massive than 1.1~$M_\odot$ (black line). This improves slightly $V_A$, while for the sample with $M_*>1.1\,M_\odot$, we find $V_A=0.68$. The gap appears more marked, and shifted to larger radius, for the metal-rich sub-sample. However, we caution that only 76 and 68 planets remain with a radius between $1.31$ and $2.38$ in these two samples; furthermore,  changing the threshold in metallicity changes the shape of the curves, so  this result should be taken with care.

\begin{figure}
        \centering
        \includegraphics[scale = 0.5]{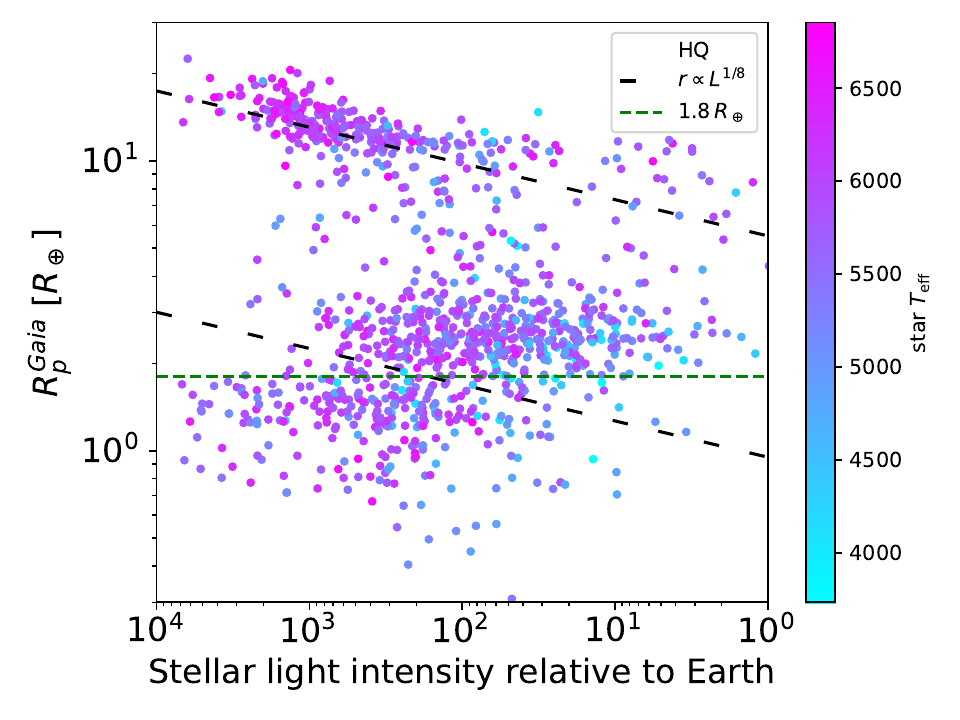}
        \includegraphics[scale = 0.5]{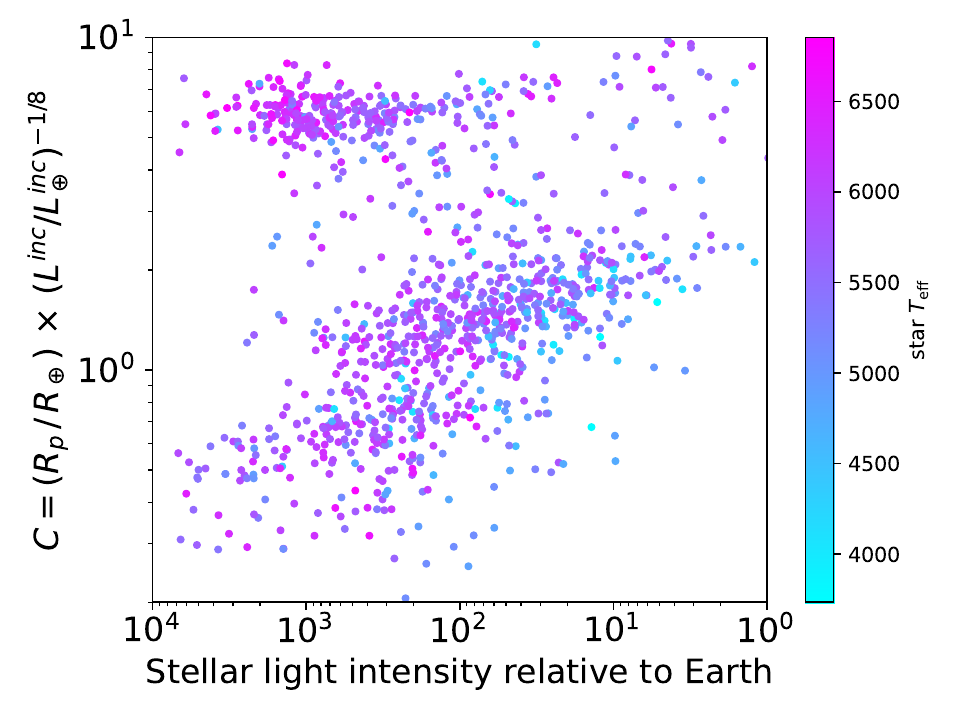}
        \caption{Top: Planet radius as a function of stellar light intensity. A separation within the group of the smaller planets appears at a radius that increases with incoming light intensity (dashed grey line). The green dashed line corresponds to the radius gap at \Rp=~1.8$R_\oplus$, as defined by \cite{Fulton17}, and the black dashed line to \Rp$=0.95~(L/L_\oplus)^{1/8}$.
        Bottom: (\Rp$/R_\oplus)/(L/L_\oplus)^{1/8}$ as a function of  stellar light intensity; the separation is now horizontal.} 
        \label{Fig:R-L}
\end{figure}

Finally, the top panel of Fig.~\ref{Fig:R-L} reproduces the right panel of Fig.~4 of \citet{Fulton18} and shows the planetary radius as a function of the incident stellar light received by the planet, 
$L_p^{inc}=\frac{L_\star}{4\pi a^2}$. The stellar light intensity is indeed known to be responsible for the evaporation of the planetary gaseous atmosphere, and hence plays a role in the distinction between Earth-like and Neptune-like planets \citep[e.g.][]{Owen2013, Lopez2013}. 
The group of small planets  (below $5~R_\oplus$) is clearly split in two blobs, which are better separated by the bent black long-dashed line defined by $R_p/R_\oplus\propto(L_p^{inc}/L_\oplus^{inc})^{1/8}$ than the green horizontal one at \Rp=1.8~$R_\oplus$. 

Hence, we show, on the $y$-axis in the bottom panel of Fig.~\ref{Fig:R-L}, the quantity
\begin{equation}
C = \left(\frac{R_p}{R_\oplus}\right) \left(\frac{L_p^{inc}}{L_\oplus^{inc}}\right)^{-1/8}.\ 
\label{eq:C}
\end{equation}
Then, the gap appears horizontal around $C\lesssim 1$ 
and the hot Jupiters population ($L^{\rm inc}>10\,L^{\rm inc}_\oplus$, $C>0.45$) has a $C$ value independent of $L^{\rm inc}$, contrary to $R_p$, which increases with $L^{\rm inc}$.

\begin{figure}
        \centering
        \includegraphics[scale = 0.5]{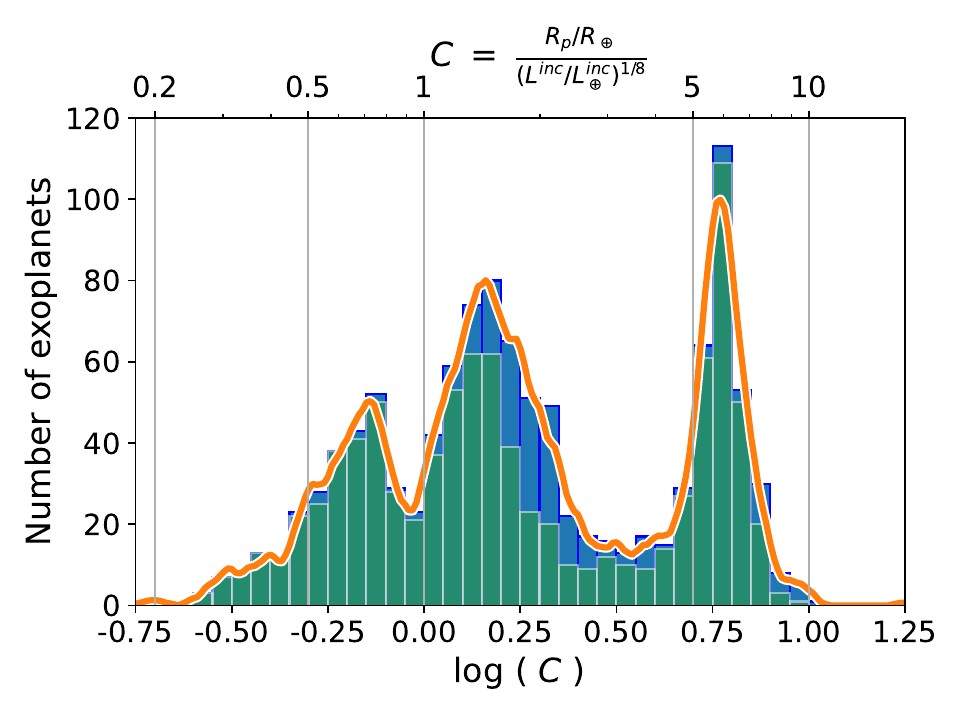}
        \includegraphics[scale = 0.5]{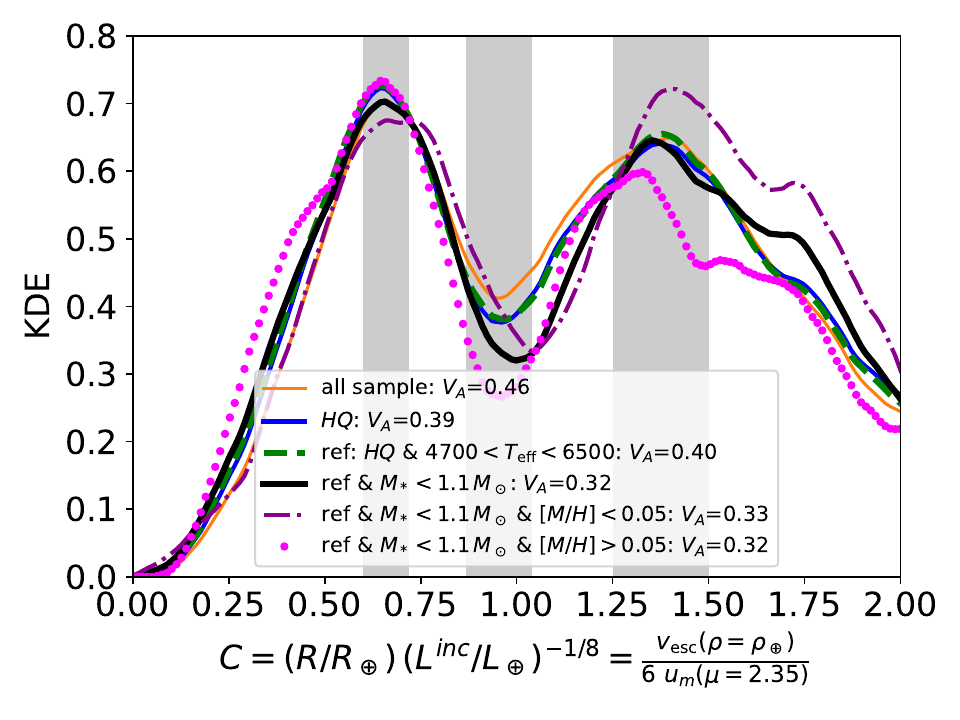}
        \caption{Top: Distribution of (\Rp$/R_\oplus)/(L/L_\oplus)^{1/8}$ in blue. Kernel Density Estimation using an Epanechnikov (parabolic) Kernel of width 0.05 dex in orange. Green shows the same as the blue histogram for the planets with $L>30L_\oplus$.
        Bottom: KDE of $C$ for non-giant planets and various sub-samples.
        }
        \label{Fig:Lagadec}
\end{figure}

Figure~\ref{Fig:Lagadec} shows in the top panel the histogram and a KDE of $\log(C)$ defined in Eq.~(\ref{eq:C}). 
We first notice that the hot Jupiters are grouped around a narrow peak with 
$\log(C)= 0.741\pm 0.096$\,dex for the population with $0.45<\log(C)<1.25$ ($C=0.741$ is the top black dashed line in Fig.~\ref{Fig:R-L}). In contrast, the peak of the hot Jupiters with 
$-0.3<\log(R_p/R_{\rm Jup})<0.5$ has 
$\log(R_p^\Gaia/R_{\rm Jup})=0.032\pm 0.121$\,dex, 
so a standard deviation 26\% larger.

The bottom panel of Fig.~\ref{Fig:Lagadec} is the KDE of $C$ for $\log(C)<0.5$ obtained with an Epanechnikov kernel of width 0.2, for different sub-samples. The shaded regions show the ranges that we have used to compute a $V_A$ value\,: [0.6,0.72], [0.867, 1.04] and [1.25, 1.5] ; they are 0.08~dex wide and equally spaced in logarithm, the middle one being centred on $C_{\rm gap}=0.95$ (the lower black dashed line in Fig.~\ref{Fig:R-L}). As expected from Fig.~\ref{Fig:R-L}, we find lower values of $V_A$ than using $R_p$ (see legend of the figure).
This argues for $C$ to be a more meaningful parameter than $R_p$ to characterise this valley.
Also, the gap appears now deeper for $M_*<1.1\,M_\odot$, independently of the metallicity of the host star, though metal-poor ones seem to host more $C>1.25$ planets than metal-rich ones.

Using $T_{\rm irr} = \left( L_p^{inc}/2\sigma\right)^{1/4}$ the equilibrium black body temperature of the irradiated hemisphere of the planet, we have the typical agitation velocity of gas molecules $u_m = \sqrt{3kT_{\rm irr}/\mu}\propto (L_p^{inc})^{1/8}$
with $k$ the Boltzmann constant and $\mu$ the mean molecular weight (2.35 g per mol for the classical H$_2$ - He composition).
In contrast, the escape velocity at the surface of a planet $v_{\rm esc} = \sqrt{2GM_p/R_p}$ is proportional to \Rp\ for a fixed density, $\rho_p$.
Thus, we have\,
\begin{equation}
    C = \frac{v_{\rm esc}}{6\,u_m}
\left(\frac{\rho}{\rho_\oplus}\frac{\mu}{2.35\,\rm g.mol^{-1}}\right)^{1/2}
\label{eq:C_vu}
,\end{equation}
so that $C$ can be seen as a measure of the effectiveness of the thermal evaporation of an H$_2$-He atmosphere;\, $C<1$ means that a non-negligible fraction of the molecules are above the escape velocity and the atmosphere eventually vanishes. All the planets below this evaporation valley would therefore have lost their primary atmosphere.

We caution however that the density, $\rho$, is not constant: there is a general trend for $\rho$ to decrease with increasing \Rp\ for planets smaller than Saturn. Thus, $v_{\rm esc}$ may have a more complicated dependence on \Rp, making this interpretation of $C$ a bit simplistic. 
Unfortunately, the planetary mass is known for too few planets around the radius valley to use the real escape velocity in this diagram.

\subsection{Particular exoplanet cases}
\begin{figure}[t]
    \centering
    \includegraphics[width=\linewidth]{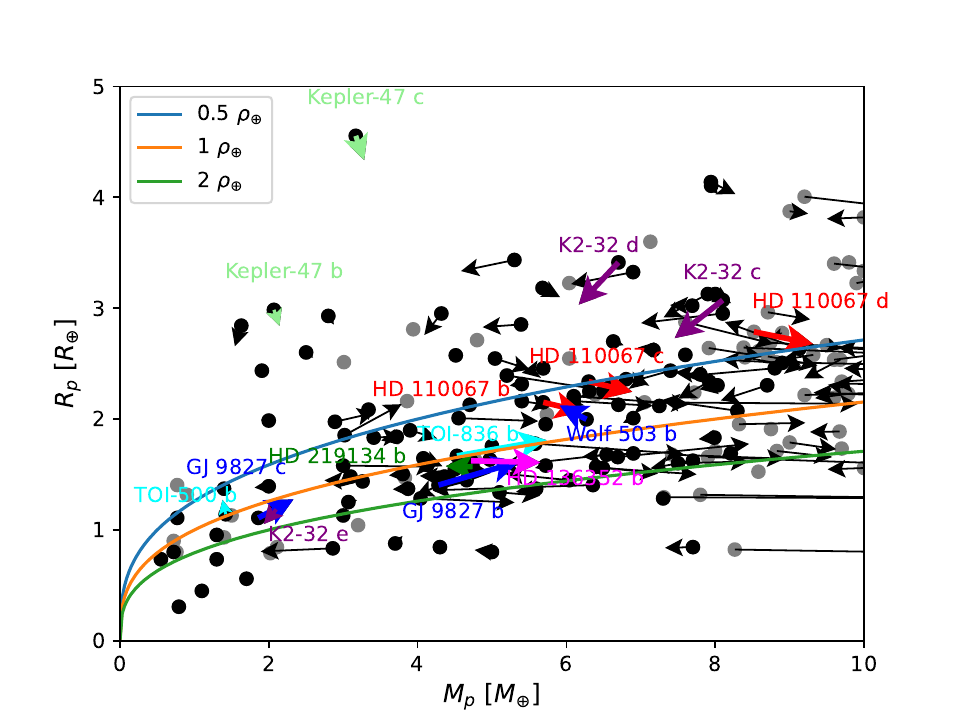}
    \caption{Variation of the position of the exoplanets in the \Rp-\Mp\ diagram caused by the different sets of stellar parameters adopted to compute them. The arrows connect the literature parameters to the \Gaia\ ones. The grey points correspond to the exoplanets for which the \Gaia\ mass is larger than $8.5 M_\oplus$. In the $HQ$ sample,  the cases for which the shift in \Mp\ or \Rp\ is higher than $1 \sigma$ are shown in colour. The coloured lines represent the iso-densities.}
    \label{fig: R-M change}
\end{figure}

When taking a more detailed look at Fig.~\ref{Fig:Mp-Rphist}, the mass and/or radius of several exoplanets differ between the literature and our \Gaia\ catalogue. Figure~\ref{fig: R-M change} illustrates the shift of the exoplanets positions in the \Mp-\Rp\ plane between both catalogues. The coloured lines represent iso-densities of 0.5, 1, and 2~Earth density. This shift is larger in mass than in radius due to the discrepancies already discussed above; we thus observe a global horizontal shift in this figure. The black points represents the exoplanets with \MpGaia<8.5~$M_\oplus$ (sub-Neptune planets), the variations for giant planets being less significant for the internal structure of the exoplanets. 
When considering the $HQ$ sample of sub-Neptune planets, we only find 15 exoplanets with a difference in \MpGaia\ or \RpGaia\ greater than $1 \sigma$. If we consider all the exoplanets (telluric and gas giants), these numbers increase\,: 230 differ from at least $1 \sigma$, 51 from $2 \sigma$ and 19 from $3 \sigma$ (365, 105, and 35, considering the full sample). Even if a $1 \sigma$ change in radius or mass can appear negligible,
it can actually lead to very different densities and thus, structures of the exoplanets, especially for the low-mass ones.
Since the relative error on the planetary parameters is larger than the one on the stellar parameters, the shifts between \Gaia\ and the literature\ masses and radius, once normalised by the $\sigma$, appear smaller for the planets than for the stars.

For instance, among the cases showing discrepancies in the $HQ$ telluric sample, we find the system of GJ 9827, that hosts 3 (known) exoplanets, and for which planets $b$ and $c$ differ of $1 \sigma$. The stellar mass found in the literature is \MstarLitt $= 0.62 \pm 0.04 M_\odot$, while \MstarGaia $= 0.87 \pm 0.23 M_\odot$. Similarly, \RstarLitt $= 0.58 \pm 0.03 R_\odot$ while \RstarGaia $= 0.67 \pm 0.03 R_\odot$. This obviously involves changes on the parameters of the exoplanets. At the end, the densities are affected, and change from 8.5, 7.5, and 2.6 g$\cdot$cm$^{-3}$ for GJ 9827 $b$, $c$, $d$ respectively, to 7.0, 6.2, and 2.1 g$\cdot$cm$^{-3}$. In Fig. \ref{fig: R-M change}, we see that the blue arrows corresponding to  GJ 9827 $b$ and $c$ cross several iso-density lines. Likewise, the arrow of HD 136352 $b$ is horizontal\,: the stellar masses are inconsistent (\MstarGaia $= 1.14\pm0.03 M_\odot$ and \MstarLitt$=0.87\pm0.04 M_\odot$), and this translates into two planetary masses of more than $1 \sigma$ difference.

In the case of TOI-500, the stellar radii vary going from the database to the \Gaia\ catalogue\,: \RstarLitt $= 0.68 \pm 0.02 R_\odot$ while \RstarGaia=0.760$\pm0.005 R_\odot$, which impacts the difference in planetary radius. Hence, the arrow on Fig. \ref{fig: R-M change} is rather vertical. Even if it seems small, we observe a decrease of 30$\%$ in the planetary density using \Gaia\ properties compared to those from the literature ('Lit.'), which suggest a huge change in the planetary composition.

Another interesting example is the super-Earth 55~Cnc~$e$. In this case, the stellar radii agree very well, but the stellar masses change from 0.91$\pm$0.01~M$_{\odot}$ (\Lit) to 0.87$\pm$0.01~M$_{\odot}$ (\Gaia). The planetary masses are consistent (\MpLitt = 8.0$\pm$0.3 and \MpGaia = 7.8$\pm$0.3~M$_\oplus$). 
However, it is interesting to note that a more direct estimation of the stellar mass was performed by \cite{Crida-etal-2018,Crida-etal-2018RNAAS}, who derived $1.015\pm 0.051~M_{\odot}$, thus a value closer to the \Lit\ one, and leading to a planetary mass of 8.59 $\pm 0.43 M_\oplus$. They find a density of $\rho_{\rm p}$=6.4~g$\cdot$cm$^{-3}$ (with \Rp=1.95$\pm0.04~R_\oplus$, a higher value than in \Lit\ and \Gaia), while $\rho_{\rm p}^{\Gaia}$=7.0~g$\cdot$cm$^{-3}$ and $\rho_{\rm p}^{\Lit}$=7.2~g$\cdot$cm$^{-3}$. 
These results have been obtained combining interferometry (to measure the stellar radius) and transit light curves,  allowing us to measure the stellar density and, thus, the stellar mass. This same method has been applied to HD~219134 \citep{Ligi2019}, for which the stellar mass is compatible with \MstarGaia\ ($\sim$ 0.72 \Mstar), but they are both smaller than in the literature. The interferometric radius is however smaller ($0.73$ \Rstar) than \RstarGaia\ and \RstarLitt\ ($\sim 0.78$ \Rstar). The other case is HD~97658, for which \cite{Ellis2021} find a mass of $0.85 \pm 0.08$ \Mstar\ from interferometry and transit light curve, while the \Gaia\ mass is $0.70 \pm 0.04$ \Mstar. The \Gaia\ radius is also higher ($0.76 \pm 0.004$ \Rstar\ versus $0.73 \pm 0.01$ \Rstar). At the end, this translates into a higher planetary density from the interferometric study. 

This method is a very good way to obtain stars benchmarking our results. The SPICA/CHARA instrument \citep{Mourard2022SPIE} should indeed offer this possibility, by combining the six telescopes of the CHARA array in a fibred instrument. For instance, the Interferometric Survey of Stellar Parameters (ISSP) survey is partly dedicated to the measurement of 
EHS radii, providing about 50 benchmark EHSs spread over the HR diagram \citep{Mourard2022SPIE}. Among these stars, 44 are in the \Gaia\ spectroscopic catalogue and 39 are in the $HQ$ sample. This makes it possible to combine both spectroscopic and interferometric high quality data, promising a very precise and accurate parametrisation.

To conclude, we can say that the variation of stellar parameters do not impact the overall population of exoplanets. However, it has an influence when looking at some peculiar exoplanets, above all if one wants to derive their composition. Generally, in our context, stellar parameters are obtained using models, but models do not take into account the external error. Thus, we can consider that a given set of internal structures obtained for an exoplanet only represent one limited solution for it.
Actually, the solutions are in fact much more numerous, i.e. the final degeneracy is larger than found in the literature.

\subsection{Exoplanets and Galactic properties of their host stars}
\begin{figure}
        \centering
        \includegraphics[scale = 0.58]{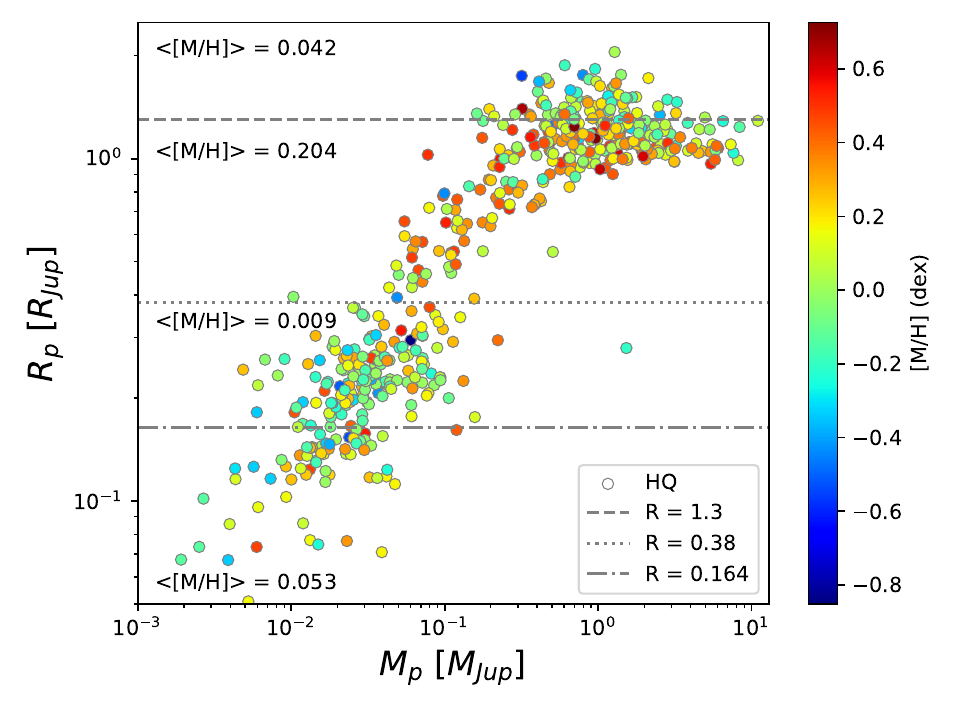}
        \caption{\Mp\ - \Rp\ for the \Gaia\ $HQ$ sample. The colour bar illustrates the metallicity of the corresponding host star. The two top grey dashed lines show the planetary radii where the average host star metallicity seems to change; the lowest one shows the radius gap. The mean metallicity in each regions is indicated (see text for more details).}
        \label{Fig:Mp-R_Meta}
\end{figure}

\begin{figure*}
        \centering
        \begin{tabular}{cc}
        \includegraphics[scale = 0.4]{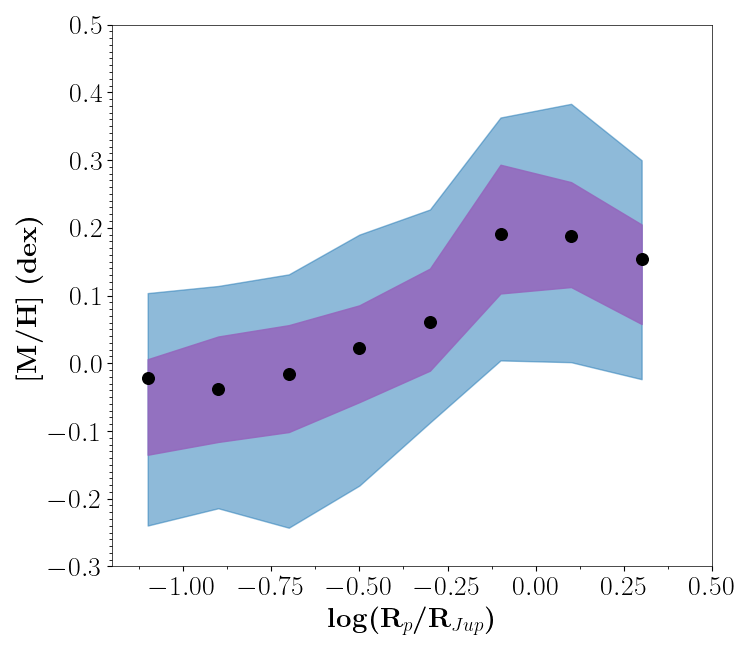} & \includegraphics[scale = 0.4]{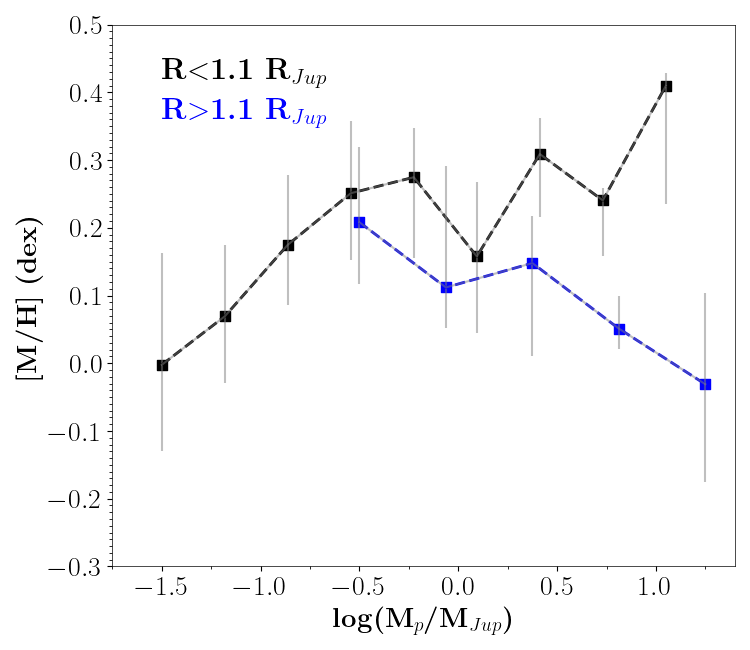}\\
        \end{tabular}
        \caption{Left panel: Median metallicity as a function of planet radius (black points). The shaded areas represent the dispersion between the 16$\%$ and the 84$\%$ (blue), and the  dispersion between the 33$\%$ and the 67$\%$ (purple). Only EHSs with an error in metallicity lower than 0.025~dex have been considered. 
        Right panel: Median metallicity as a function of planet mass for planets with a radius lower (black) and higher (blue) than 1.1 R$_{Jup}$. Error bars show the 33$\%$-67$\%$ interval of the metallicity distribution for each bin in planetary radius.}
        \label{Fig:MetRM}
\end{figure*}

\begin{figure*}[h!]
        \centering
        \begin{tabular}{cc}
         \includegraphics[scale = 0.4]{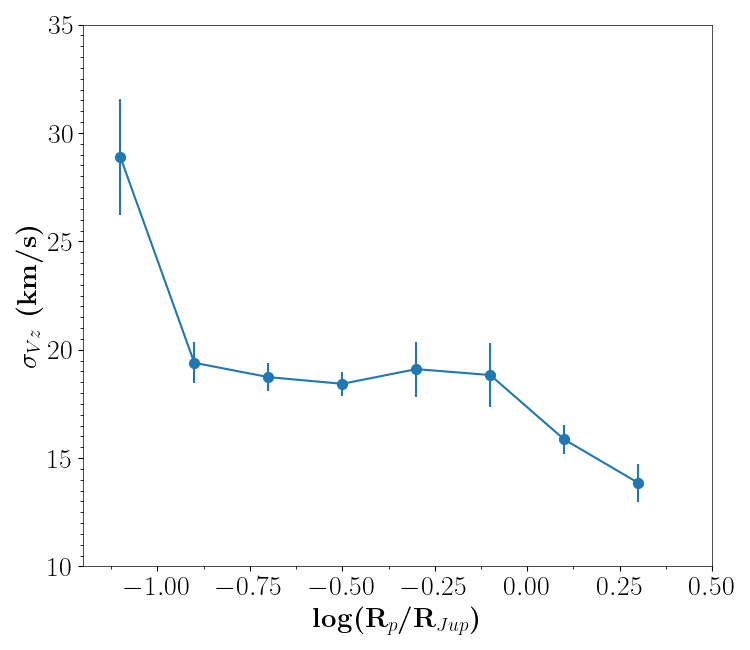}    &  \includegraphics[scale = 0.4]{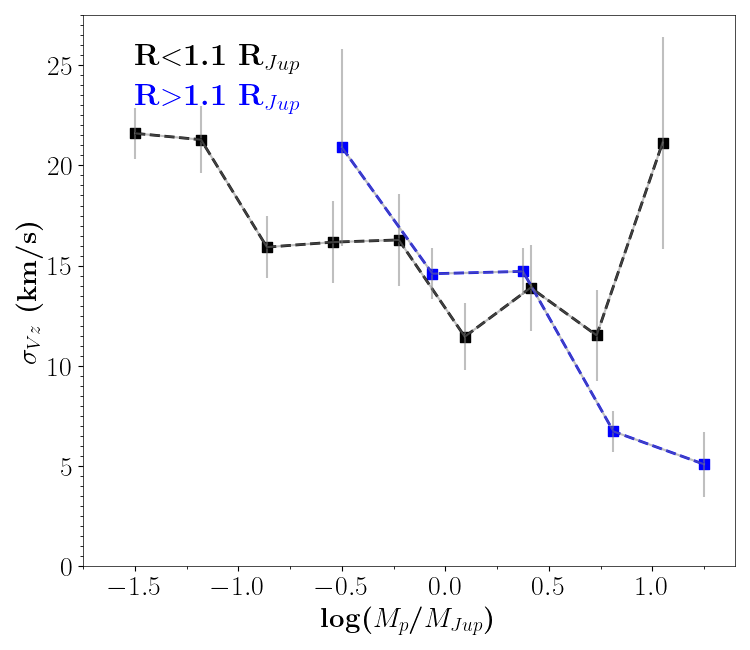}\\
        \end{tabular}
        \caption{Vertical Galactic velocity dispersion as a function of planet radius (left panel) and planet mass (right panel). As in Fig.~\ref{Fig:MetRM}, two regimes of planet radius are shown in the right panel. Error bars show the uncertainty in the standard deviation estimated from $\sqrt{\sigma^2/2(n-1)}$.
        }
        \label{Fig:vzpl}
\end{figure*}

\begin{figure*}
        \centering
        \begin{tabular}{cc}
        \includegraphics[scale = 0.5]{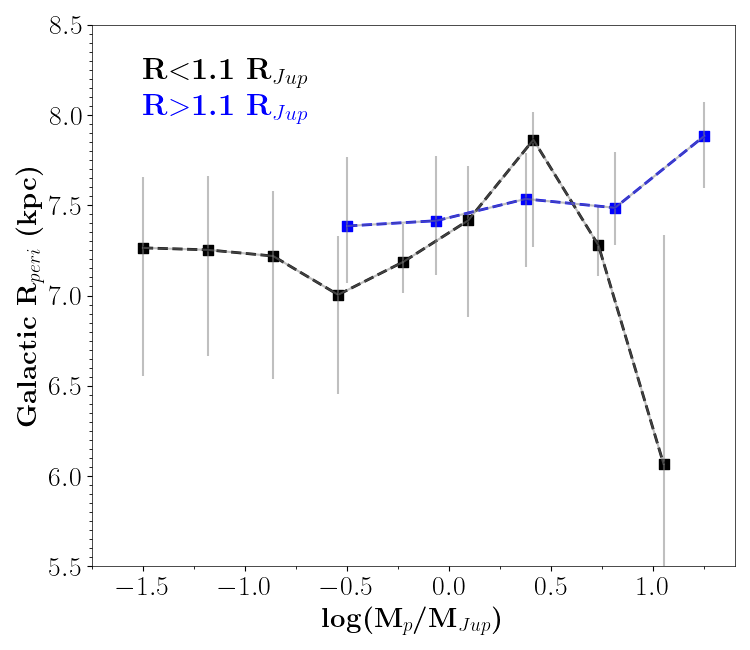} &  \includegraphics[scale = 0.5]{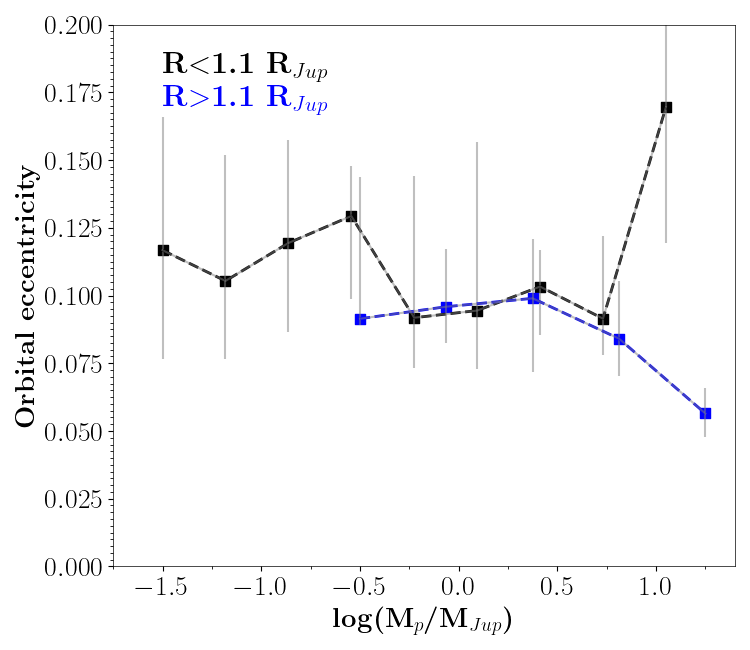}\\
        \end{tabular}
        \caption{Left panel: Median Galactic pericentre of host star as a function of planet mass. Right panel: Median eccentricity of the host star's Galactic orbit as a function of planet mass. Error bars show the 33$\%$-67$\%$ interval of the R$_{peri}$ and eccentricity distributions for each bin in planetary mass. }
        \label{Fig:Rguide}
\end{figure*}

As discussed in Sect.~\ref{Sec:MassRadDist}, the EHS chemo-physical properties are known to influence the characteristics of their planetary system. Conversely, the stellar properties are contingent upon its affiliation with a Galactic population (e.g. young thin disc, old thick disc, halo, etc). In this section, we discuss the studied exoplanet sample from a Galactic perspective, particularly focusing on (i) the chemical composition, (ii) the age, using the velocity dispersion as a proxy, and (iii) the Galactic orbital parameters of the EHSs. 

It is important to recall that the history of formation and evolution of the Milky Way imposes that the three above mentioned parameters are complexly related \cite[see, for instance][]{PVP_Ale}. The main factors governing Galactic trends are the star formation rate (depending on the density and the turbulence of the gas), the chemical enrichment of the interstellar medium by the stellar nucleosynthesis, and the dynamical evolution of the disc (influenced by the spiral arms and the central bar) through processes as gravitational scattering of stars and angular-momentum transport. As a consequence, global chrono-chemo-kinematical trends appear in the Milky Way cartography (c.f. Sect~\ref{Sec:EHSpopulation}). For instance, the stellar metallicity increases towards the internal regions of the Galaxy (with generally higher star formation rates) and for younger stars (further enriched by previous stellar generations). However, the recent advances on the characterisation of Galactic populations show a high degree of dispersion in those general trends, as expected from the above mentioned evolutionary processes (e.g. orbital heating and radial migration) and from non-axisymmetric disc features (e.g. spiral arms). Interestingly, the Sun itself is thought to have migrated from more internal Galactic regions \citep[e.g.][and references there in]{Baba2024}.
Moreover, the Milky Way is not a closed system, and its history is marked by the accretion of extragalactic stars and gas (from satellite mergers or isolated gas infalls). Therefore, the evolution of its chemical and dynamical characteristics is discontinuous over time. 
In this context,  the question arises as to whether an isotropic distribution of exoplanet properties in the Milky Way is to be expected. Furthermore, we consider whether the current sample of exoplanets, which is limited to the Solar neighbourhood, is suitable for deriving intrinsic trends in planetary properties. The following analysis  is aimed at addressing those two challenging questions by adopting, as done all along this article, only the new planetary masses and radii derived in the present work.

Figure~\ref{Fig:Mp-R_Meta} displays the \Mp --\Rp\ distribution for the \Gaia\ $HQ$ sample, colour-coded with the host star metallicity. 
In agreement with the literature, we recover a general trend of higher metallicity of the EHS for higher planetary masses.
However, the variation of [M/H] appears to be more directly correlated to \Rp. The horizontal grey lines separate the planets in four bins in radius. The bottom line corresponds to the so-called radius valley \citep[see][and sect.~\ref{sec:Fulton}]{Fulton17}, and does not seem to separate two populations of EHSs with significantly different metallicities. The middle line separates super-Earths or mini-Neptunes from Giant planets and splits the population of EHSs into two samples with quite different [M/H]. Finally, the top line goes through the population of hot Jupiters, and shows that the more inflated ones seem to be hosted by less metallic stars. The mean metallicity in each bin is shown on the figure, and the standard deviation is 0.23~dex in all four samples. Only the bin 0.38\Rjup<\Rp<1.3\Rjup\ stands out of the others, but only by $\sim0.16$ dex, that is within less than 1$\sigma$.

To complement this analysis, we have explored the median metallicity distribution of the EHSs as a function of the logarithm of planet radius or mass, as presented in Fig.~\ref{Fig:MetRM}. The left panel of the figure shows a clear dependence of the median planet radius with the metallicity, with larger planets having on average a more metal-rich host star, as already illustrated in Fig.~\ref{Fig:Mp-R_Meta}. 
The quite smooth increasing behaviour suggests a continuous dependence, particularly for small planets defined as 
$\log($\Rp$/$\Rjup$) < -0.25$. 

As shown in the right panel of Fig. \ref{Fig:MetRM} the metallicity distribution as a function of planet mass is bimodal. An increase of the median metallicity with planet mass is observed in the low mass regime ($\log(M_p/M_{Jup})\lesssim-0.6$). For higher masses, two opposite behaviours are observed: while denser planets (R $\lesssim$ 1.1 R$_{Jup}$) tend to be more massive as the metallicity of the host star increases, inflated planets are more massive for less metallic hosts. As a consequence, inflated planets tend to disappear for host stars whose metallicity is higher than about 1.5 times Solar (\meta$\ga$0.2~dex). On the contrary, giant denser planets are still present in this super-metal-rich regime. This bimodal behaviour of giant planets explains the flattening of the [M/H] versus $\log($\Rp$/$\Rjup$)$ curve for higher planet radii. 
Interestingly enough, this separation between dense and inflated giant planets in different metallicity regimes seems to be intrinsic to the planetary physics. 
It had been proposed that hot Jupiters around metal-rich stars tend to be more metal-rich and therefore smaller than around metal-poor stars \citep{Tristan06, Claire13}, which would, at least qualitatively, account for the bimodality seen in Fig.~\ref{Fig:MetRM}. This trend however has not been confirmed from a sample of 24 cool giant planet systems \citep{Teske19}. Our analysis calls for a re-examination of this important problem linked to planet formation.
Moreover, as discussed later, this bimodality also imposes important constraints on the diversity of exoplanets in the Galactic ecosystem.

\smallskip

To better understand the link between the above presented trends and the underlying Galactic populations, we have explored the Galactic kinematic properties.  The stellar velocity dispersion in the Galaxy is known to increase with time \citep[e.g.][and references therein]{Wielen75,Hayden18,Hayden20}. This progressive dynamical heating of disc stars can be explained by the diffusion of the orbits with time due to interactions with local fluctuations of the gravitational field \citep[e.g.][]{Wielen77}. As a consequence, the Galactic velocity dispersion can be used as an age proxy of the corresponding stellar population. Recently, \cite{Swastik23}, using \Gaia\ DR3 data and \gspspec\ chemical information, concluded that stellar systems with small planets and those with giant planets (as defined from the planetary mass) differ in key aspects of Galactic space velocity and stellar orbital parameters.  In particular, the hosts of giant planetary systems present a lower Galactic velocity dispersion compared to the population of stars harbouring small planets. The observed velocity dispersion difference is consistent with an age difference measured for a sub-sample of the stars by the same authors, with the hosts of giant planets being younger.

We have therefore investigated whether the above described results are confirmed by our catalogue. Our database of exoplanets has a larger proportion of giant planets than the \cite{Swastik23} one. 
Figure~\ref{Fig:vzpl} presents the vertical Galactic velocity dispersion, $\sigma_{Vz}$, as a function of planet radius (left panel) and mass (right panel). A trend between the planetary radius and $\sigma_{Vz}$ is observed, as already noticed by \cite{Vardan13, Swastik23, Unni2022}. 
However, a more complex picture emerges when both the radius and the mass are taken into account.
The distribution of velocity dispersion as a
function of planet mass (right panel of Fig.~\ref{Fig:vzpl}), confirms the bi-modality observed  in Fig.~\ref{Fig:MetRM}. For masses higher than around 0.25 M$_{Jup}$ ($\log(M_p/M_{Jup})\sim-0.6$), denser and inflated planets present opposite behaviours. The less metallic stars  hosting very massive inflated planets show a low Galactic velocity dispersion typical of a young stellar populations. On the contrary, the more metal-rich host stars of massive denser planets present higher $\sigma_{Vz}$ values typical of older populations. 
This result seems in contradiction with the expected outcome of Galactic chemical evolution, which increases the stellar metallicity with time. However, as explained before, the chemical enrichment pace depends on the star formation rate, which was probably higher in the internal Galactic regions, causing a negative metallicity radial gradient in the disc. The observed behaviour could therefore be the result of the radial mixing of stars in the disc.
To verify this hypothesis, we have studied the properties of the EHS Galactic orbits as a function of planet mass. Figure~\ref{Fig:Rguide} shows the median value of the Galactic orbital pericentre of the EHSs  (R$_{peri}$, left panel) and the orbital eccentricity (right panel) as a function of $\log(M_p/M_{Jup})$. While the exoplanet sample is currently located in the Solar neighbourhood, the orbits of their host stars reach more internal regions of the Galaxy. Once again, the dichotomy between inflated and dense giant planets is observed in the regime of very high masses. For inflated planets, a slight increase in R$_{peri}$ with planet mass and a decrease in the Galactic orbital eccentricity is observed. In contrast, massive dense planets are present around stars with more eccentric orbits reaching the Solar neighbourhood from more internal Galactic regions. This is consistent with their larger vertical velocity dispersion and higher metallicities, suggesting that dense giant planets are preferentially found around older stars born in the inner Milky Way disc, where inflated planets tend to be absent. 

In conclusion, we have found a metallicity difference between giant dense and inflated planets that seems to be intrinsic to the planetary physics. If this dichotomy is confirmed, it implies that the diversity of giant exoplanets depends on their Galactic birth locus, with dense giant planets more numerous than inflated ones in Galactic areas, where the metallicity is higher than Solar; for instance, as seen in the central regions of the Milky Way.

\section{Conclusion}
\label{Sec:Conclu}
We have built a new large homogeneous catalogue of 3556 exoplanets and 2573 exoplanet host stars properties, called the \Gaia\ spectroscopic catalogue. The stellar atmospheric and chemical properties, namely, metallicity and enrichment in $\alpha$-elements with respect to iron, come from the 
\Gaia\  \gspspec\ spectroscopic stellar parameters catalogue, complemented with  data from ground-based spectroscopic surveys. Then, adopting \Gaia\ astrometry and photometry, we have  estimated the EHS luminosity, radius, and mass without adopting any priors from stellar evolution models. The Galactic positions, velocities, and orbital
parameters were also computed. 
The selection of exoplanets was made from the EES and NEA databases and their radii and masses were then consistently derived from the stellar parameters with additional information collected in these databases.
These planetary properties are thus fully consistent with the stellar ones. A system of quality flag was finally implemented in order to select the highest-quality data to be used in association with the reported uncertainties for every stellar and planetary parameters.

The newly derived EHS properties are first compared to literature values. It is found that our effective temperatures, stellar luminosities, and radii   agree rather well with the literature values,
but they are more precise. For instance, \RstarGaia\ 
 are derived with typically less than $3\%$ uncertainty; namely, more than a factor two better than in the literature. As a consequence, the uncertainty on the planetary radii can be significantly reduced, in particular for aptly observed transit light curves, the stellar radii often being the limiting factor for a precise planetary radius determination in the case of transiting exoplanets.
On the contrary, the \Gaia\ catalogue and database stellar masses are found to differ more, 
leading to possible large external errors that are usually
not considered. For consistency and homogeneous reasons, we therefore recommend to adopt the  \Gaia\ spectroscopic EHS catalogue, in particular for its high-quality sub-sample. 

The population of EHSs is found to be very diverse, either in its stellar atmospheric parameters, chemical and/or Galactic properties.
About 85\%  are dwarf main sequence stars and have close to Solar mass (or slightly lower), with the metallicity and \AF\ varying in the [-0.7, +0.5]~dex and [-0.2, +0.3]~dex] ranges, respectively. On the contrary, a few other stars reach luminosity as high as $\sim$1000~$L_\odot$ with a radius of $\sim$100~$R_\odot$ and a metallicity of around -0.5~dex with \AF$\sim$+0.2~dex. 

From a Galactic point of view, and as expected, all the exoplanets detected so far are located in the Solar vicinity or in the $Kepler$
field, close to the Local spiral arm. 
The closest exoplanets from the Sun
orbit around giant stars characterised by rather high (Solar and slightly higher) metallicities, typical of the Solar local volume. 
The EHS \AF\ versus \meta\ trend is characteristic of the Galactic disc,
with larger enhancement in $\alpha$-elements at lower metallicity. Moreover, from their \AF, location close to the Galactic plane,  kinematics, and circular orbits, most EHSs
belong to the thin disc.
However, several other EHSs are found at higher
distances from the plane with rather eccentric orbits,  and
low-metallicity, typical of the thick disc or halo.
Finally, we found a metallicity difference between giant dense and inflated planets that seems to be intrinsic to the planetary physics. This would imply that the diversity of giant exoplanets depends on their Galactic birth locus. Dense giant planets would be more numerous than inflated ones in Galactic areas with super Solar metallicities as, for instance, in the Milky Way central regions.

Regarding the planetary mass and radius, most values in the literature and our catalogue 
agree rather well. 
We however observe a larger discrepancy for planetary masses than for planetary radii. This is partly explained by the fact that the stellar mass is one of the most difficult parameter to estimate. We recall here that we adopted strict criteria when deriving \MstarGaia\ that, actually, led to the rejection of several masses. On the contrary, literature stellar masses are expected to be quite heterogeneous since they come from different works and techniques.
In any case, we retrieved the classical \Mp -\Rp\ distribution, characterised by 
a general trend of
higher EHS metallicity for more massive planets. We also clearly see a continuous trend of more metal-rich and \AF -poor planets having larger radii.
However, even if small, the discrepancies in planetary properties can lead to different internal structures; in particular, for telluric or Neptune-like planets. That is why taking into account the external errors on parameters is important. We have indeed identified a few cases where the discrepancy between \Gaia\ and the literature\ exoplanetary parameters can be important and likely  lead to very different compositions. 

The radius valley is retrieved at $\sim$1.8~$R_{\rm \oplus}$ but it is more easily identified in the \Gaia\ spectroscopic catalogue, probably because of its more precise stellar hence planetary radii. A clear trend of increasing gap radius with increasing \meta\ is also identified in the [$\sim$1.7, $\sim$2.0]~$R_\oplus$ range.
Finally, examining the stellar light illuminating the exoplanet atmosphere, we can see that the
small planets group actually consists of two sub-groups, with a separation found at lower radii for larger
illumination. We also propose that considering the effectiveness of the thermal evaporation of the planetary atmospheres instead of the stellar luminosity could help to interpret this separation.
\\

In the future, it is expected that the number of parametrised stars by the \gspspec\ module
will strongly increase with forthcoming \Gaia\ data releases. Moreover, the parameter quality will be also improved thanks to the increasing \SNR\ of the \Gaia/RVS spectra. This will obviously impact the parameter quality of the exoplanets. Therefore, future \Gaia\ spectroscopic catalogues of EHS and exoplanet properties should be even more complete,
homogeneous, and precise. 
In this context, the Milky Way exoplanetary content  needs to be more widely explored and, in particular, over the different Galactic populations far from the present local sample. Detections of new exoplanets outside the Solar volume would indeed be very valuable for planetary formation scenarios. We note that the EHS properties of these future detections are probably already present in the DR3 \gspspec\ catalogue or will be published in the next releases.
We also point out that the present work could already be extended to the EHS (and, thus, planetary) chemical properties, thanks to the several individual chemical abundances already published within the \gspspec\ catalogue. In this Milky Way context, the PLATO ESA mission should provide significant advances \citep[see e.g.][]{PLATO24}.
Finally, a future complementary project could consist of a homogeneous re-analysis of all the available archive high-resolution spectra ($R\ga 10^5$) of EHSs; in particular, to complete and more deeply explore the \gspspec\ individual abundances as a function of the exoplanet properties. 

\section{Data availability}
Tables 1 and 2 are only available in electronic form at the CDS via anonymous ftp to cdsarc.u-strasbg.fr (130.79.128.5) or via http://cdsweb.u-strasbg.fr/cgi-bin/qcat?J/A+A/.

\begin{acknowledgements}
This work has made use of data from the European Space Agency (ESA)
mission \Gaia\ (https://www.cosmos.esa.int/gaia), processed by the \Gaia\ Data Processing and Analysis Consortium (DPAC, https://www.cosmos.esa.int/web/gaia/dpac/consortium). Funding for the DPAC has been provided by national institutions, in particular the institutions participating in the Gaia Multilateral Agreement.\\

PdL and ARB acknowledge partial funding from the European
Union’s Horizon 2020 research and innovation program under SPACE-H2020
grant agreement number 101004214 (EXPLORE project). 
RL was partially supported by the European Research Council (ERC) under the European Union’s Horizon 2020 research and innovation programme (Grant agreement No. 101019653) and by DFG-ANR supported GEPARD project (ANR-18-CE92-0044 DFG: KL 650/31-1).
RL and AC thank G. Kordopatis for useful discussions about catalogue queries. We thank E. Poggio for providing material for Fig.~\ref{Fig:EHSinMW}. We also thank T. Guillot, A. Petit and D. Valencia for helpful discussions. Finally, the referee is thanked for his fruitful comments.\\

We used the IPython package \citep{ipython}, NumPy \citep{NumPy}, Matplotlib \citep{Matplotlib}, Pandas, TOPCAT \citep{Topcat} and the SIMBAD database,
operated at CDS, Strasbourg, France \citep{Simbad}. 
We also used data obtained from or tools provided by the portal exoplanet.eu of The Extrasolar
Planets Encyclopaedia. This research has made use of the NASA Exoplanet Archive, which is operated by the California Institute of Technology, under contract with the National Aeronautics and Space Administration under the Exoplanet Exploration Program.
\end{acknowledgements}

\bibliographystyle{aa}
\bibliography{ref}

\begin{appendix}

\section{Additional validation of the derived radii and masses}
\label{Appendix:Valid}

\begin{figure}[h]
    \centering    
    \includegraphics[width=0.42\textwidth]{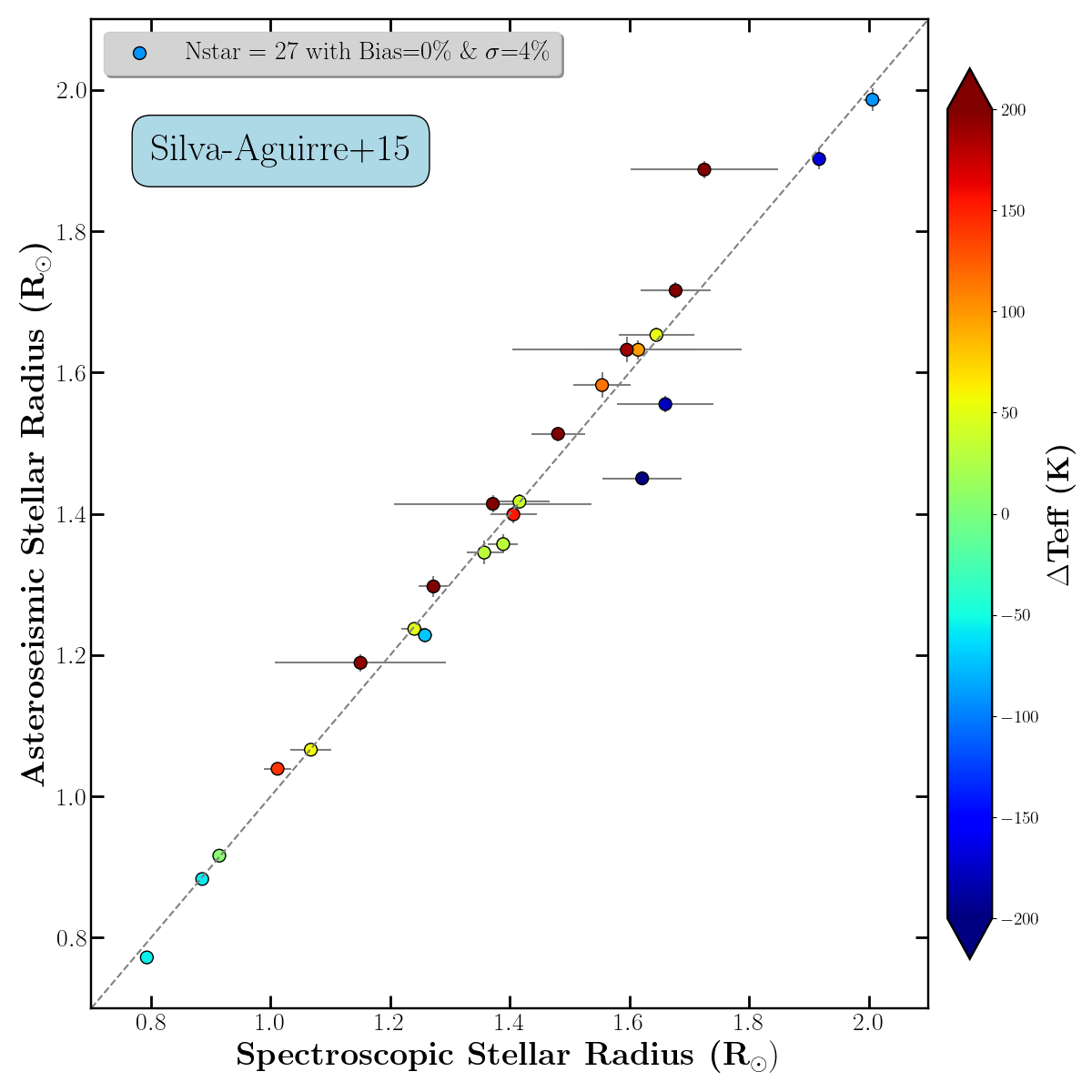} 
    \includegraphics[width=0.42\textwidth]{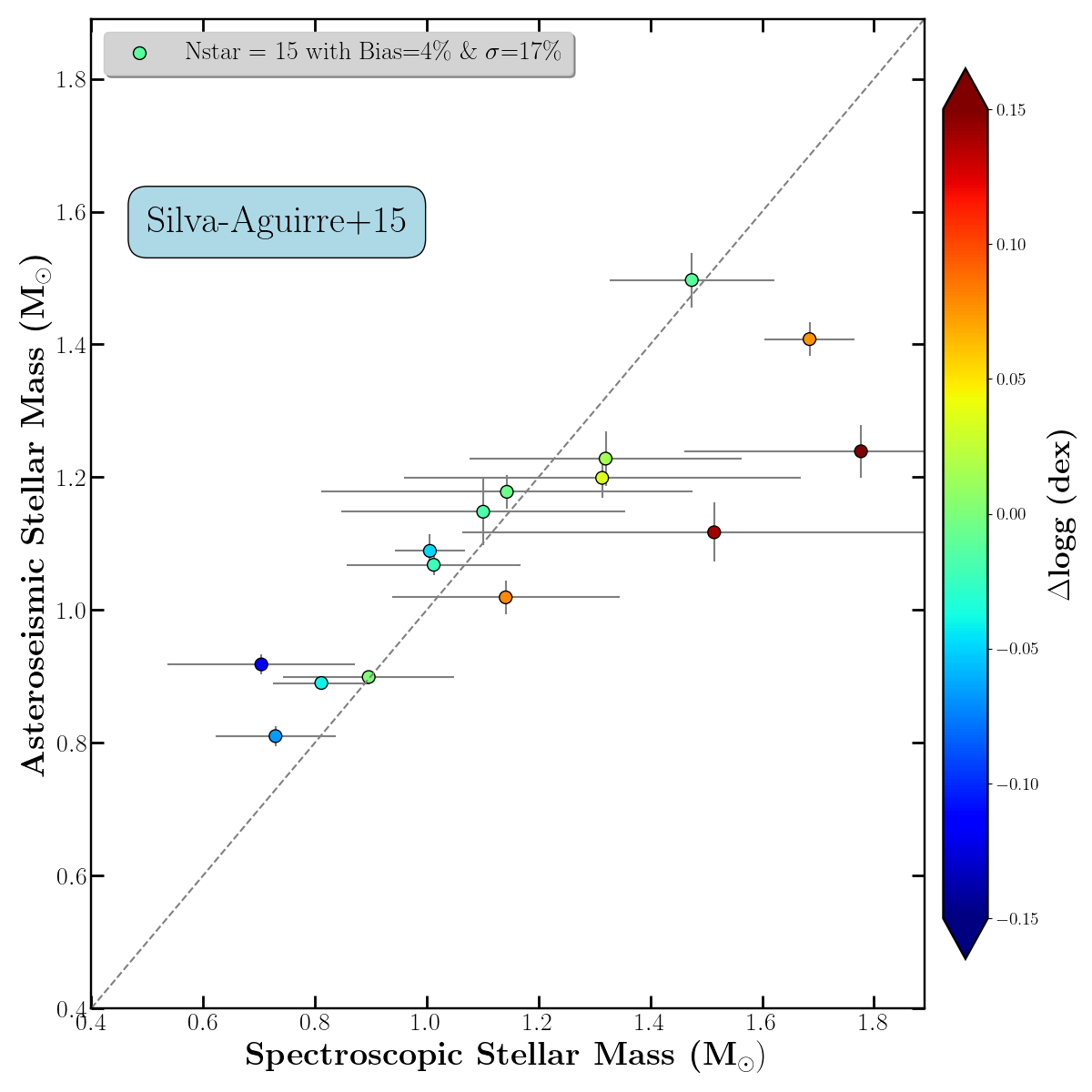}     
    \caption{Comparison between the \gspspec\ spectroscopic stellar radii and masses (left and right panels, respectively) and those derived thanks to asteroseismic data by \citet{Silva15}. The 1:1 relation is shown as a dashed grey line. The colour-codes represent the difference in \T\ and \g\ between both studies.
    The mean  $R_\star^\Gaia$ and  $M_\star^\Gaia$ relative differences and associated standard deviations indicated in each panel are in the sense of (Astero - \gspspec)/Astero.} 
    \label{Fig:ValidMR}
\end{figure}

To complement the radius and mass validations presented in Sect.~\ref{Sec:LMR}, we added a specific comparison for 
small sub-samples of EHSs with radius and mass derived from asteroseismic data. Indeed, thanks to $Kepler$ and additional ground-based spectroscopic data, \cite{Silva15} derived fundamental properties of 33 EHSs\footnote{A more detailed validation of stellar radii and masses based on the complete sample (not only EHSs) of \cite{Silva15} can be found in \cite{Recio24}.}. Among them, 27 stars are found in our EHS sample with \gspspec\ parameterisation and excellent astrometric data (\Gaia\ renormalised unit weight error, {\it ruwe}, factor less than 1.4), allowing us to directly compare their properties. 
First, it can be noted that, for these EHSs in common, \cite{Silva15} adopted \T, \g, \meta, distances and luminosities that agree well within error bars (which can be quite large for some EHSs with low \SNR\ spectra) with our adopted values. The mean and standard deviation of the differences for these five parameters are indeed (50, 194~K), (0.04, 0.17~dex), (0.0, 0.13~dex), (-1.0, 19~pc) and (0.05, 0.18~$L_\odot$), respectively. We show in  Fig.~\ref{Fig:ValidMR} the comparison between both sets of  radii and masses. For the radii comparison (left panel),
the agreement is excellent with no bias and a standard deviation of $\sim$0.06~R$_\odot$, i.e. $\sim$4\% in the sense of ($R_\star^{\rm Astero}$ - $R_\star^\Gaia$)/$R_\star^{\rm Astero}$. Moreover, it can be seen that the three stars that most depart from the 1:1 relation are those with the largest differences in \T. Since stellar radii are directly derived
from 
the effective temperature (see Eq.~\ref{eq:Lstar}), it is natural that studies adopting different \T\ have derived different $R_\star$.
If we had filtered out such stars with large \T\ differences, the agreement would have been even tighter. 
Regarding the stellar masses, 
larger discrepancies are expected when larger \T\ and/or \g\ differences exist between the compared studies (see Eq.~\ref{eq:Mstar}). We therefore compare the stellar masses in the right panel of Fig.~\ref{Fig:ValidMR} for stars having $\Delta$\T<250~K and $\Delta$\g<0.25~dex and relative errors in mass less than 50\%. Fifteen stars satisfy this selection criteria. It can be seen that the agreement for the masses is very satisfactory (bias of 4\% only, in the sense of ($M_\star^{\rm Astero}$ - $M_\star^\Gaia$)/$M_\star^{\rm Astero}$), although the mass uncertainties are larger for the spectroscopic than the asteroseismic study. The associated standard deviation (17\%) is also rather small. As expected, the largest differences in mass are found for the largest differences in \g, and the estimated bias can be fully explained by the bias in \g\ reported above. On the contrary, the stars with the best agreement on \g\ (light green point) have a spectroscopic and asteroseismic mass in excellent agreement.

\section{Complementary effective temperature - luminosity diagrams}
\label{Appendix:L-T}
The data associated with Fig.~\ref{Fig:L_Teff1} confirm that all EHSs are correctly located along the main sequence and the giant branches in the   $L_\star^\Gaia$- \T\ plane, confirming their high-quality parameters. 
\begin{figure*}[]
    \centering
    \includegraphics[width=0.48\textwidth]{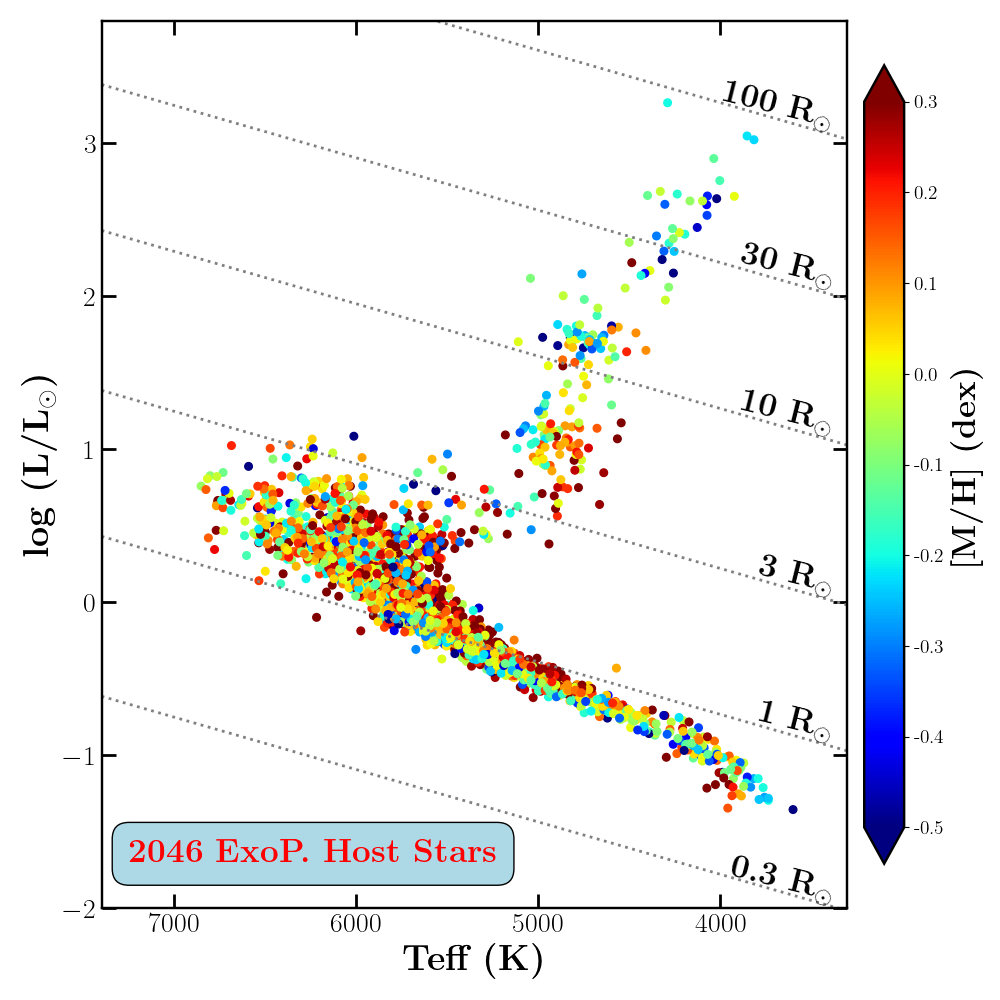} 
    \includegraphics[width=0.48\textwidth]{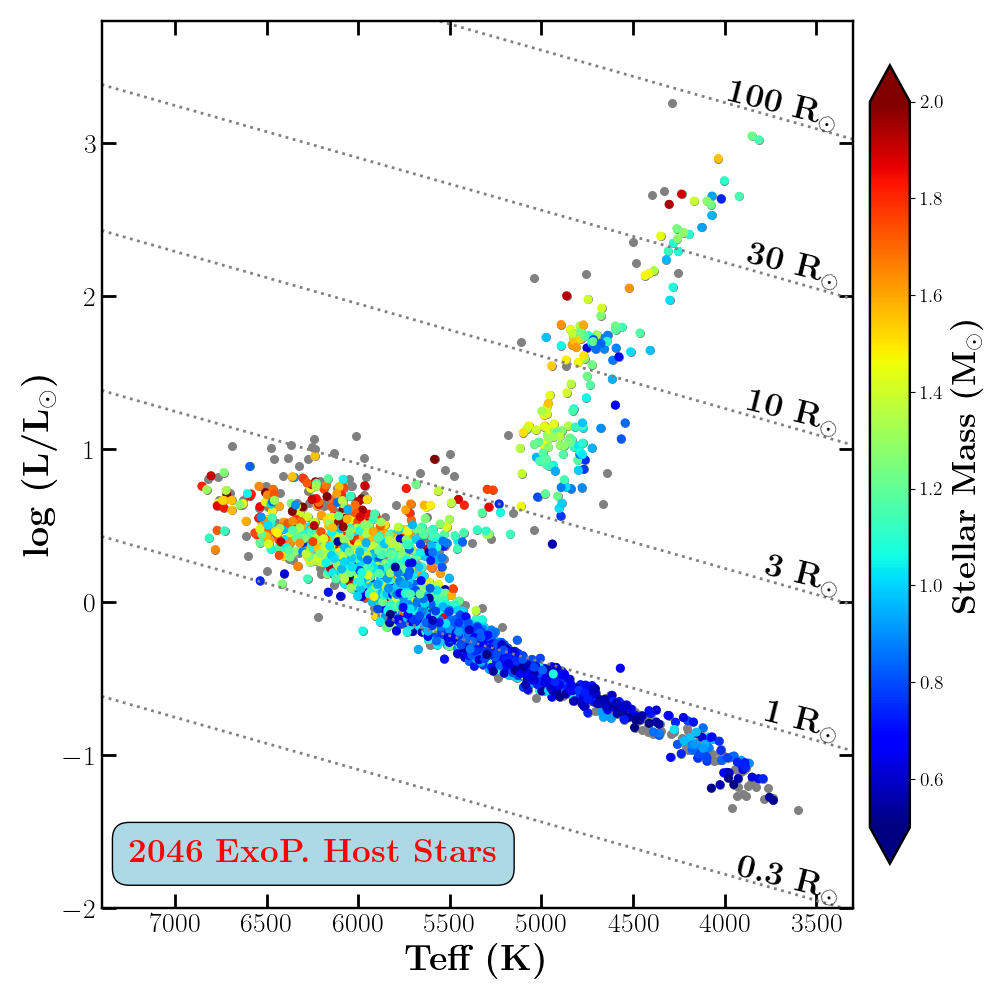} 
    \caption{Same as Fig. \ref{Fig:L_Teff1} but colour-coded with the metallicity and stellar mass (right and left panels, respectively). The EHSs without published mass are plotted in grey in the right panel.} 
    \label{Fig:L_Teff2}
\end{figure*}

\section{EHS kinematics}
\label{Appendix:kinematics}
\begin{figure}[]
        \centering
        \includegraphics[scale = 0.25]{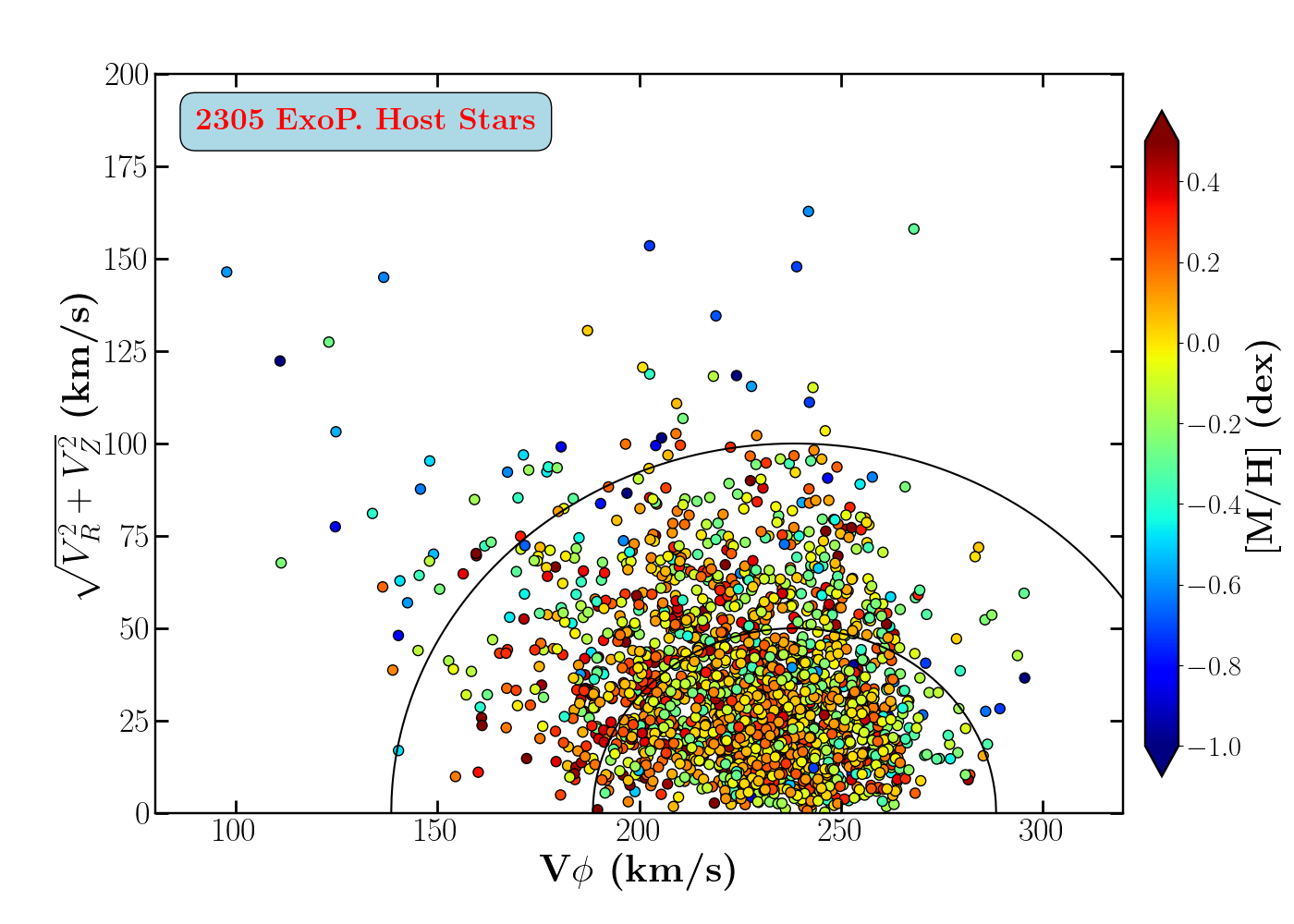}
        \caption{Toomre diagram of the EHSs, colour-coded with their metallicity. The lines correspond to $V_{\rm (Tot,~LSR)}$ equal to 50 and 100~km.s$^{-1}$, respectively. Stars having $V_{\rm (Tot,~LSR)}$ smaller than $\sim$50~km.s$^{-1}$ are expected to belong to the Galactic thin disc.}
        \label{Fig:Toomre}
\end{figure}
 The kinematic properties of these EHSs in a Toomre diagram, colour-coded with their metallicity are shown in Fig.~\ref{Fig:Toomre}. We recall that the rotational velocity in the Galactic plane ($V_\phi$) has a typical value for thin-disc stars in the Solar vicinity around $\sim$240~km.s$^{-1}$. 
In this figure, the dotted lines represent the total velocities respective to the LSR:
$V_{\rm (Tot,~LSR)} = \sqrt{V_R^2 + V_Z^2 + (V_\Phi-V_{\rm LSR})^2}$ equal to 50 and 100~km.s$^{-1}$.
The regions below the first circular dotted line are
those where thin disc stars are preferably found \cite[see e.g.][]{PVP_Ale}. From Fig.~\ref{Fig:MetaAlpha}, \ref{Fig:RZ}, \ref{Fig:Rperi}, and  \ref{Fig:Toomre}, it can be seen that most EHSs actually belong to the Galactic thin disc: they are close to the Galactic plane with  kinematics and orbital properties typical of thin disc stars: rather circular orbits, $V_\phi \sim 220-260$~km.s$^{-1}$ and $V_{\rm (Tot,~LSR)}\la$50~km.s$^{-1}$. However, several other stars have velocities revealing their Galactic thick disc or halo nature, that is also confirmed by their lower metallicity.
We also note that a similar diagram colour-coded with the eccentricity does not reveal any peculiarity: EHSs with the most eccentric orbits are those belonging to the Galactic thick disc or halo.
 
\end{appendix}

\end{document}